\documentclass[amsfonts, amssymb, amsmath, prl,twocolumn,showkeys, floatfix, twoside]{revtex4-2}
\usepackage[english]{babel}
\usepackage[utf8]{inputenc}
\usepackage[colorinlistoftodos, color=green!40, prependcaption]{todonotes}
\usepackage{amsthm}
\usepackage{mathtools}
\usepackage{physics}
\usepackage{xcolor}
\usepackage{graphicx}
\usepackage[left=23mm,right=13mm,top=35mm,columnsep=15pt]{geometry} 
\usepackage{adjustbox}
\usepackage{placeins}
\usepackage[T1]{fontenc}
\usepackage{lipsum}
\usepackage{csquotes}
\usepackage{soul}
\usepackage{xcolor}
\usepackage{bm}
\usepackage[pdftex, pdftitle={Article}, pdfauthor={Author}]{hyperref} 
\begin{document}
\title{Dynamical geometric modes in non-Euclidean plates}

\author{Joseph C. Roback$^{1}$}
\email{joseph.roback@colorado.edu}
\thanks{These authors contributed equally to this work.}
\author{Carlos E. Moguel-Lehmer$^{2}$}
\email{cmoguell@syr.edu}
\thanks{These authors contributed equally to this work.}
\author{Katharina A. Fransen$^{1}$}
\email{Katharina.Fransen@colorado.edu}
\thanks{These authors contributed equally to this work.}
\author{Christian D. Santangelo$^{2}$}
\email{cdsantan@syr.edu}
\author{Ryan C. Hayward$^{1,3}$}
\email{ryan.hayward@colorado.edu}

\affiliation{$^{1}$Department of Chemical and Biological Engineering, University of Colorado Boulder, Boulder, 80303, Colorado, USA.}
\affiliation{$^{2}$Department of Physics, Syracuse University, Syracuse, New York 13244, USA}
\affiliation{$^{3}$Department of Chemical and Biological Engineering, University of  Colorado Boulder, Boulder, 80303, Colorado, USA.}

\begin{abstract}
When subjected to specific prestresses, continuum elastic shells can exhibit geometric zero modes: complex motions that require vanishing elastic energy to excite, enabling them to be driven by weak and generic energy inputs. Despite recent interest in these modes, we understand very little about their dynamical properties. Non-Euclidean plates modeled on minimal surfaces are one example in which prestresses and geometry combine to produce a continuum of ground states that the plate can explore through a geometric zero mode. We demonstrate that a non-Euclidean plate with metric corresponding to Enneper's minimal surface exhibits the predicted continuous stability, but this degeneracy is ultimately lifted by aging. Despite developing a preferred configuration, the zero mode remains the softest mode. Using a combination of analytical theory and experiments, we show that the elastodynamics of this soft mode is captured by the dynamics of a damped pendulum. A periodic driving uncovers resonance phenomena in this pendulum mode, such as small oscillations and steady rotations, but mixes with an additional flapping mode at high frequencies.

\end{abstract}


\maketitle

There is a vast literature exploring the origin and consequences of zero modes in discrete elastic systems, including ordered and disordered lattices \cite{kane2014topological, lubensky2015phonons, rocklin2016mechanical}, spring networks and vertex models \cite{damavandi2022energetic, huang2022shear, hernandez2022anomalous}, and mechanical metamaterials \cite{coulais2016combinatorial, bertoldi2017flexible, rocklin2017transformable, hu2023engineering, dudek2025shape}.
On the other hand, continuum zero modes, sometimes called vanishing-stiffness modes \cite{chibbaro2022chaotic, guest2011zero, hamouche2017multi, schenk2014zero, seffen2011prestressed}, geometric zero modes \cite{baumann2018motorizing}, or zero elastic energy modes \cite{deng2024light, zang2025self}, remain challenging because they give rise to geometrically complex motions that are poorly captured by linear elasticity. Experiments have observed rolling in rings \cite{baumann2018motorizing, zhao2023self}, climbing and everting in tori, knots, and twisted ribbons \cite{zhu2024animating, deng2024light, nie2021light}, and ``spinning'' in saddles \cite{holmes2011bending, chibbaro2022chaotic}. These examples speak to the role of complex geometries in enabling zero modes, highlighting a need to develop geometrical models of the large-scale dynamics.

Incompatible elasticity \cite{efrati2009elastic} provides a powerful framework for realizing prestressed sheets that exhibit ultrasoft deformations \cite{levin2016anomalously}.
In non-Euclidean plates, for example, in-plane stresses arise due to a spatial variation of swelling or shrinkage of the sheet's area that leads to out-of-plane buckling without a corresponding preference for a local curvature. When the resulting film adopts the shape of a minimal surface -- a surface having zero mean curvature at every point -- it can exhibit a geometry-induced zero mode, the Bonnet isometry, which arises from an internal symmetry of the sheet as opposed to material symmetries such as homogeneity and isotropy.

This work explores the dynamics of this geometric zero mode in a special class of non-Euclidean plates modeled on the Enneper minimal surface with two lobes \cite{kim2012designing}. The zero mode takes the form of an apparent ``spinning'' of the lobes without a coinciding rotation of the material, similar to what is observed in Refs. \cite{holmes2011bending, chibbaro2022chaotic}.
After fabrication, Enneper non-Euclidean plates show continuous stability that follow the Bonnet isometry, but as they age they develop a preferred configuration. We show that the softest mode of the aged system can still be described geometrically by the Bonnet isometry and that the resulting dynamics are well-described by a mapping to a damped pendulum. Periodic perturbations can drive the system from
irregular oscillations to a continuous rotation-like mode, but at high driving frequencies in fresh samples, the simple dynamics is interrupted by hybridization with a ``flapping'' excitation.

\begin{figure}[h!]
    \centering
    \includegraphics[width=1\linewidth]{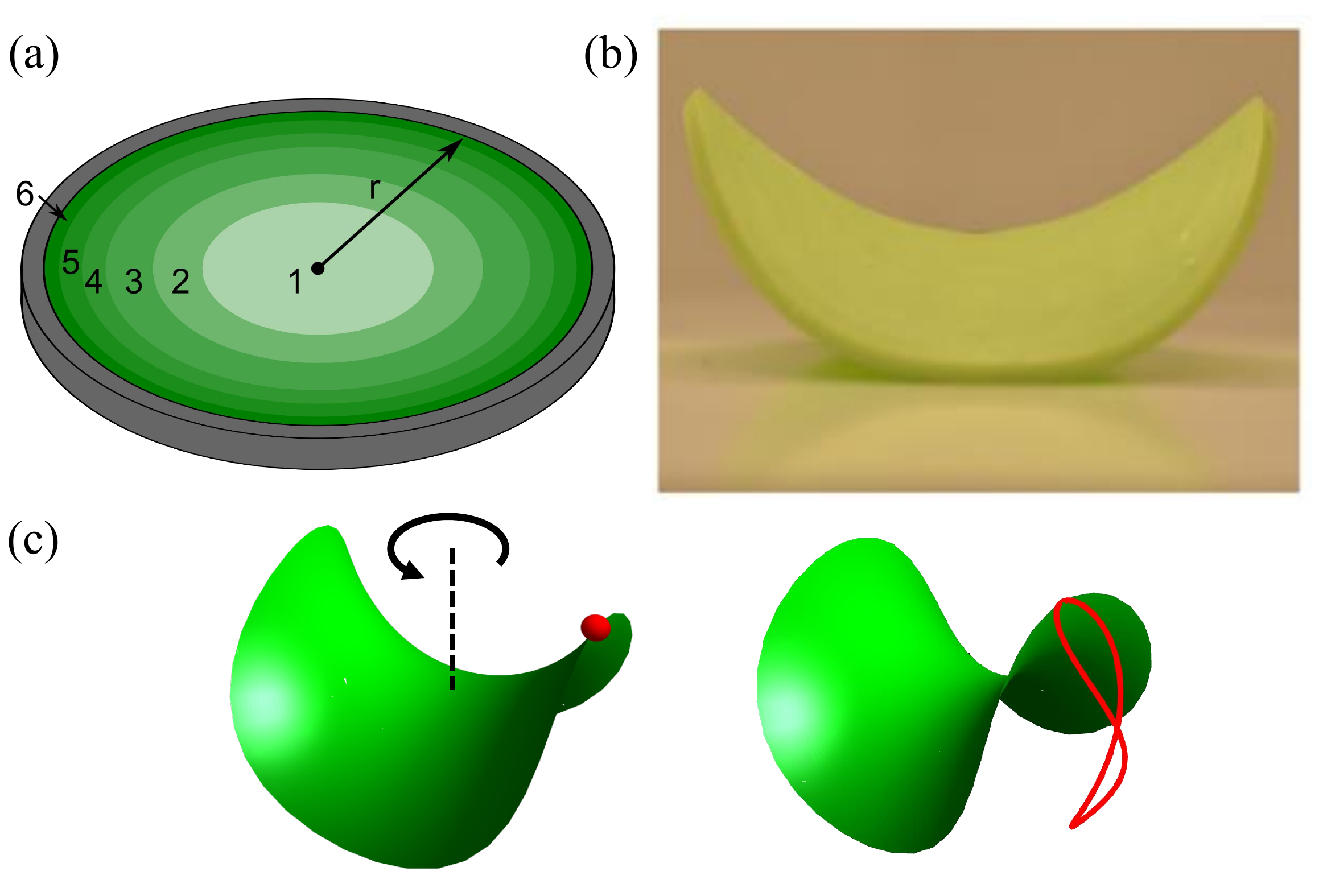}
    \caption{(a) A metric is programmed by inducing an in-plane radial shrinkage on a flat plate. Each annulus had a different degree of shrinkage, forcing the plate to buckle out of plane. (b) Buckled non-Euclidean plate with the shape of Enneper's minimal surface with two lobes. The lobes' orientation is spontaneously chosen by breaking the starting radial symmetry. (c) Zero mode in Enneper non-Euclidean plates. While it appears to be a rotation of the sheet, it is the directions of the principal curvatures that rotate. The red curve is the trajectory of a material point (red dot) under the Bonnet isometry after removing rigid-body motions, indicating that the sheet is not undergoing a rigid-body rotation.}
    \label{fig 1}
\end{figure}

Minimal surfaces are a special class of surfaces whose mean curvature, $H = (c_+ + c_-)/2 = 0$, where the functions $c_\pm$ are the principal curvatures, i.e. the largest and smallest curvature, at each point, but whose Gaussian curvature $K = c_+ c_- < 0$ everywhere \cite{do2016differential}.
Such surfaces naturally admit a deformation mode, the Bonnet isometry, that changes their shape without stretching or changing its mean curvature, preserving the elastic energy during deformation.
Intuitively, this deformation is associated with a broken symmetry: minimal surfaces are locally saddle-shaped, spontaneously breaking the isotropy of the material. The Bonnet isometry corresponds to rotating the orientation of the raised and lowered portions of the surface, though the global deformation can be significantly more complex (see \cite{fogden1999continuous, levin2016anomalously, moguel2025topological, sun2021fractional} and examples in \cite{dierkes2010minimal}).

A non-Euclidean plate shaped like Enneper's minimal surface can be produced by programming an areal shrinking $\Omega(r) = c (1+r^2/R^2)^2$ at every point, where $c$ is a proportionality constant. \cite{kim2012designing} (Fig. 1a). Though the areal shrinking is axisymmetric, the resulting surface is globally saddle-like (Fig. 1b). This spontaneously broken symmetry determines the form of the Bonnet isometry in this case: although it could be mistaken for a simple rigid body rotation, it is not the surface itself that rotates, but rather the directions of its principal curvatures (Fig. 1c). 


We fabricated surfaces from a commercial polyvinylsiloxane (PVS) resin (Zhermack Elite Double 32), which we diluted with silicone oil prior to curing (see SI, Sec 1). After curing, the material was soaked in solvent to remove the silicone oil, resulting in volumetric shrinkage. We introduced an in-plane gradient in shrinkage magnitude by varying the amount of added silicone oil in an annulus-by-annulus approach, allowing for the prescription of curved metrics. We approximated $\Omega(r)$ by a step function representing six annuli of constant shrinkage. After curing all six layers, washing, and drying, the targeted Enneper surface was formed (\emph{Materials and Methods}). 

We measured the orientation angle of the direction of the maximum principal curvature relative to the horizontal in the image (\emph{Materials and Methods} for more details). The horizontal axis of the image was chosen as a reference point for ease of data processing. Immediately after fabrication, the surfaces were continuously stable; that is, there was no preference for lobe direction, and manual manipulation of the lobes to a different orientation resulted in them remaining indefinitely (Fig. 2a; SI Movie 1). This behavior is indicative of a zero-energy mode.

\begin{figure}[h!]
    \centering
    \includegraphics[width=1\linewidth]{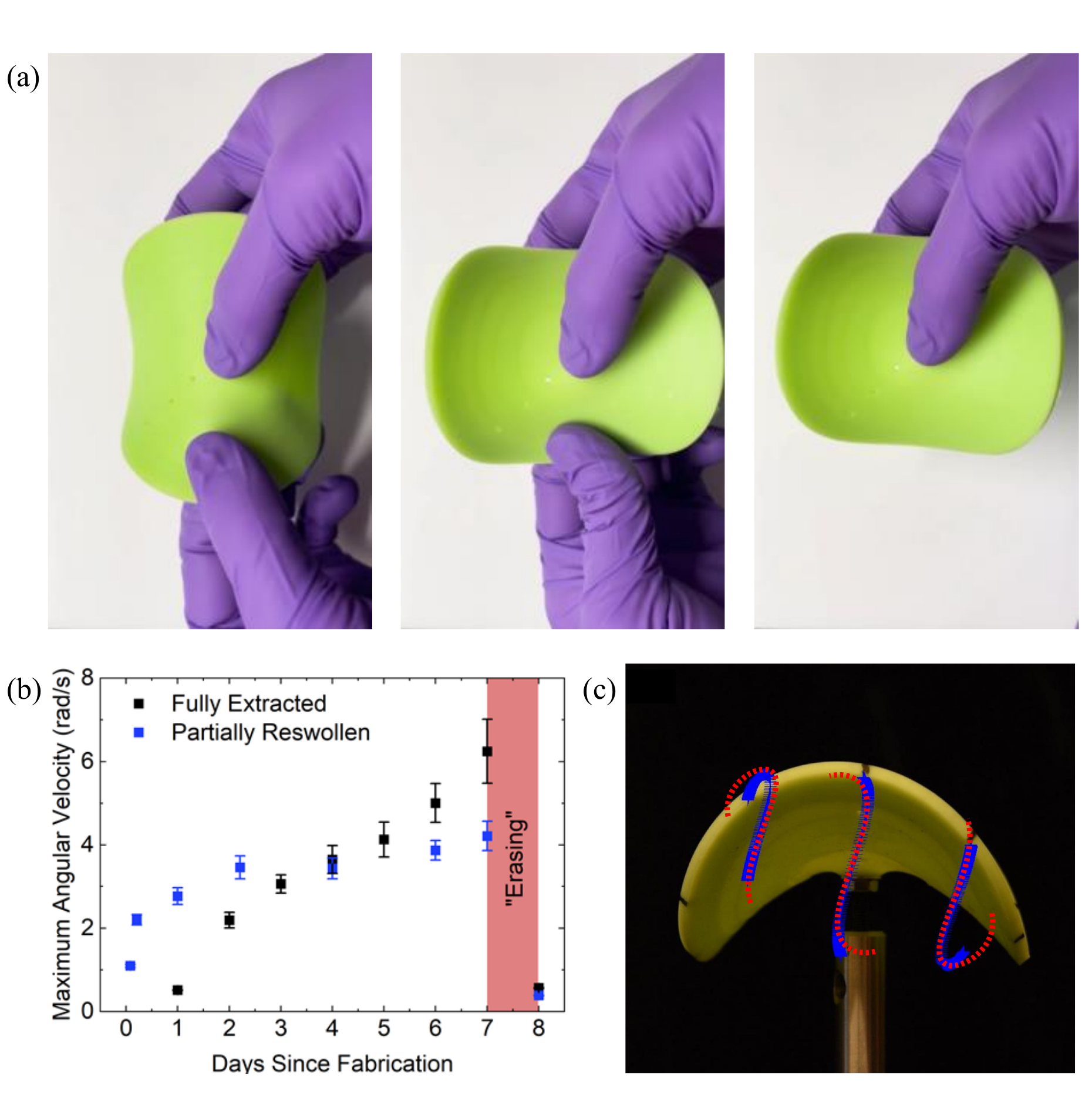}
    \caption{(a) Snapshots of a continuously stable non-Euclidean Enneper plate after fabrication. b) Maximum angular velocity of the direction of the maximum principal curvature relative to the horizontal as a function of time after changing the lobes' orientation. The initial velocity was small, consistent with the continuous stability, but increased as aging occurred. Fixing the lobes' orientation dramatically decreased the velocity by erasing the preferred configuration. c) Comparison between edge trajectories in experiments (solid blue) and Eq.~(\ref{associate family}) (dotted red), showing that the aged sample deformed along the Bonnet isometry.}
    
    \label{fig 3}
\end{figure}


Leaving the samples undisturbed for a few hours to one day resulted in the surfaces adopting a preferred configuration. Subsequently, manually forcing the lobes to a different orientation and releasing them resulted in rotation back to the preferred configuration (SI Movie 2). We observed an increase in rotation rate as the samples were left undisturbed for longer. To characterize this behavior, videos were recorded of the surfaces returning to the preferred configuration after manual forcing, and the maximum angular velocity, defined as the steepest slope of the measured angle, as a function of time after fabrication was measured (\emph{Materials and Methods}). After an initial period of continuous stability, the maximum angular velocity of the Bonnet mode slowly increased over time. After one week, the lobes were manually rotated to $\pi/2$ from their preferred orientation and the surface was taped in that configuration for 24 hours, restoring the continuous stability. Repeating the same experiment resulted in a significant reduction in the maximum velocity (Fig. 2b). However, regardless of how long the surface was fixed in the non-preferred orientation (up to two weeks was tested), it never became the preferred one. 

We attribute the aging into a preferred configuration to small amounts of residual uncrosslinked polymer chains diffusing through the material, partially relaxing the prestress inherent to a non-Euclidean plate and thereby defining some amount of preferred curvature. To test this hypothesis, the surface was reswollen in a 0.1 vol \% solution of silicone oil in hexanes, then dried to remove the hexanes. This process resulted in additional uncrosslinked silicone oil being introduced to the sample and increased its weight by approximately 1\%. The maximum angular velocity increased faster and plateaued at a lower value compared to the fully extracted sample, indicating that the introduced silicone oil accelerated the aging. Restoring the continuous stability significantly reduced the maximum angular velocity, consistent with the results from the fully extracted sample (Fig. 2b). This data provides strong support for the proposed explanation that the diffusion of residual uncrosslinked material is responsible for the aging behavior.

To explore the shape of the deformation, we parameterize the Bonnet isometry with an angle $\theta$ according to \cite{colding2011course, dierkes2010minimal, nitsche1989lectures}
\begin{equation}
    \mathbf{X}\left(r,\phi; \theta\right) = \mathcal{R}(\theta, \hat{\mathbf{N}}) \left[ \cos \theta \  \mathbf{X}_0\left(r,\phi\right) + \sin \theta \  \mathbf{X}_C\left(r,\phi\right) \right],
\label{associate family}
\end{equation}
where $\mathcal{R}(\theta,\hat{\mathbf{N}})$ is a matrix rotating the disk by an angle $\theta$ about the unit normal vector $\hat{\mathbf{N}}$ at its center, $\mathbf{X}_C(r, \phi)$ is the conjugate minimal surface of $\mathbf{X}_0$, defined by the condition $\mathbf{X}\left(r,\phi; \pi/2\right) = \mathbf{X}_C(r, \phi)$, and $\mathbf{X}(r, \phi; \theta)$ is a minimal surface for all values of $\theta$. The rotation matrix is necessary because the experimental setup fixed the disk's center and prevented rigid-body rotations, a boundary condition that is not reflected in the bare Bonnet isometry. This choice fixes the radial tangent vector at the origin such that material points on the surface describe figure-eight paths instead of the elliptical orbits of the bare isometry (see SI Sec. 2.2 and \emph{Materials and Methods}). 

The points along the edge of the surface were tracked during relaxation back to the preferred configuration. We found that the material points adopted the expected figure-eight path as the principal curvature directions rotated. This behavior was confirmed by comparing with the trajectories of material points according to Eq. (\ref{associate family}) (Fig. 2c; SI Movie 3). 
The good agreement between the experimental and theoretical trajectories shows that the deformation is well approximated by the Bonnet isometry.

To model the dynamics of the sheet after aging, we assume the plate
develops a prescribed second fundamental form, $\bar{b}_{ij}\left(\bar{\theta}\right)$, with the form of a member of the associate family for a specific Bonnet angle. A short computation, using ${b}_{ij}\left(\theta + \bar{\theta}\right)$ as the induced second fundamental form, shows that the bending energy is (see SI Sec.~2.3)
\begin{equation}
    \mathcal{E}_b = \frac{h^3 \bar{k}}{3} \left[\int \mathrm{dA} \ \left(-\mathrm{K}\right)\right] \left(1 - \cos \theta\right)
\label{Shell Bending energy},
\end{equation}
where $\mathrm{dA} = \mathrm{dr} \mathrm{ d\phi} \sqrt{\det \bar{g}}$ is the prescribed area measure, $\mathrm{K}$ is the Gaussian curvature, $h$ is the thickness, and $\bar{k} =\mathrm{Y}/\left[12 (1 + \nu)\right]$ is the saddle-splay bending rigidity for a homogeneous isotropic shell with Young's modulus $\mathrm{Y}$ and Poisson's ratio $\nu$ \cite{efrati2009elastic}. Eq.~(\ref{Shell Bending energy}) shows that there is an energetic cost associated with deviations of the induced configuration from the preferred configuration, which is characterized by the deformation angle, $\theta$.

We obtain a dynamical theory by promoting 
the Bonnet angle $\theta$
to a dynamical degree of freedom, $\theta(t)$.
To model the sheet as a viscoelastic solid, we consider a three-dimensional Kelvin-Voigt (KV) model \cite{christensen2013theory, phan2013understanding}. Since the thickness is smaller than the characteristic radius of curvature of the midsurface, we employ the Kirchhoff-Love hypothesis \cite{love1944treatise} to dimensionally reduce the KV model to an effective equation of motion for $\theta(t)$ (see SI Sec.~3)
\begin{equation}
    \mathrm{I}\ddot{\theta}(t) + 2\beta \dot{\theta}(t) + h^3 \bar{k} \sin\theta(t) = 0.
\label{equation of motion pendulum}
\end{equation}
Eq.~(\ref{equation of motion pendulum}) is equivalent to the equation of motion for a damped pendulum with gravitational force \cite{llandau60:mech}, with moment of inertia $\mathrm{I} \approx c^4 h R^4 \rho  + \mathcal{O}\left[\left(h/\mathrm{R}\right)^2\right]$, where $\rho$ is the density of the material and $\mathrm{R}$ is the radius of curvature of the sheet, and damping coefficient $\beta = h^3\tau$, where $\tau$ is the shear viscosity. Assuming uniform mass density across the sheet, this is a long and light pendulum of length $l \propto \mathrm{R}^2/h^3$ and mass $m \propto (h^3/\mathrm{R})^2$.

When the deformation is small, Eq.~(\ref{equation of motion pendulum}) predicts the Bonnet angle performs damped harmonic motion with relaxation time $t_\text{relax} = \beta / \mathrm{I}$, natural frequency $f_0 = (1/2\pi)\sqrt{h^3 \bar{k}/\mathrm{I}}$, and characteristic angular velocity $\omega_0 = 2\pi f_0$. The measured orientation angle from the sheet's free dynamics corroborates this, showing a transition from overdamped to underdamped behavior $96$ hours after fabrication.  We estimated $t_\text{relax} \approx 1/6$ s and $f_0 \approx 6$ Hz from the underdamped limit (\emph{Materials and Methods}).

Our geometric description provides a simple explanation for the observed transition from overdamped to underdamped dynamics. We consider an effective description of the aging and assume the sheet develops the prescribed second fundamental form $\delta \bar{b}(\bar{\theta})$. This prescribed curvature interpolates between an Enneper non-Euclidean plate, at $\delta = 0$, and an Enneper shell, at $\delta = 1$. Assuming a quasistatic process, in which the aging is slower than the timescale of the mechanical deformation, we derive Eq.~(\ref{equation of motion pendulum}) with renormalized coefficients, $\bar{k} \mapsto \delta \bar{k}$ and $\mathrm{I} \mapsto \mathrm{I} + \mathrm{I}_\delta \delta^2$ where $\mathrm{I}_\delta$ is a finite quantity, but the shear viscosity $\tau$ doesn't renormalize (see SI Sec.~3.5). Thus, the natural frequency $f_0 \to 0$ as $\delta \to 0$ while the relaxation time $t_\text{relax}$ stays finite. Since underdamped oscillations require $\omega_0^2 - 1/t^2_\text{relax} > 0$, we expect overdamped dynamics for sheets close to the non-Euclidean limit and a transition to underdamped dynamics for aged sheets, in agreement with experiments.

To further test the validity of the pendulum model, we probed the sheet's dynamics by exciting the Bonnet isometry with a rotating magnetic field.  Simpler perturbations, such as vertical or horizontal oscillations of the sheet, do not excite the effective pendulum, as the Bonnet isometry does not couple to translational modes (see SI Sec.~4.1).
A small permanent magnet was glued to a point on the edge of the sample, with the center of the sample mounted on a post.  At the same height as the center point of the Enneper surface, a large permanent magnet was clamped on an electric motor with driving frequency $f$, and was positioned such that the center of the large magnet was a distance $d$ from the edge of the sample, where $d$ ranged from 4 to 7 cm (\emph{Materials and Methods}). Note that the orientation angle measured in our experiments is twice the Bonnet angle  \cite{moguel2025topological}.

\begin{figure}[h!]
    \centering
    \includegraphics[width=1\linewidth]{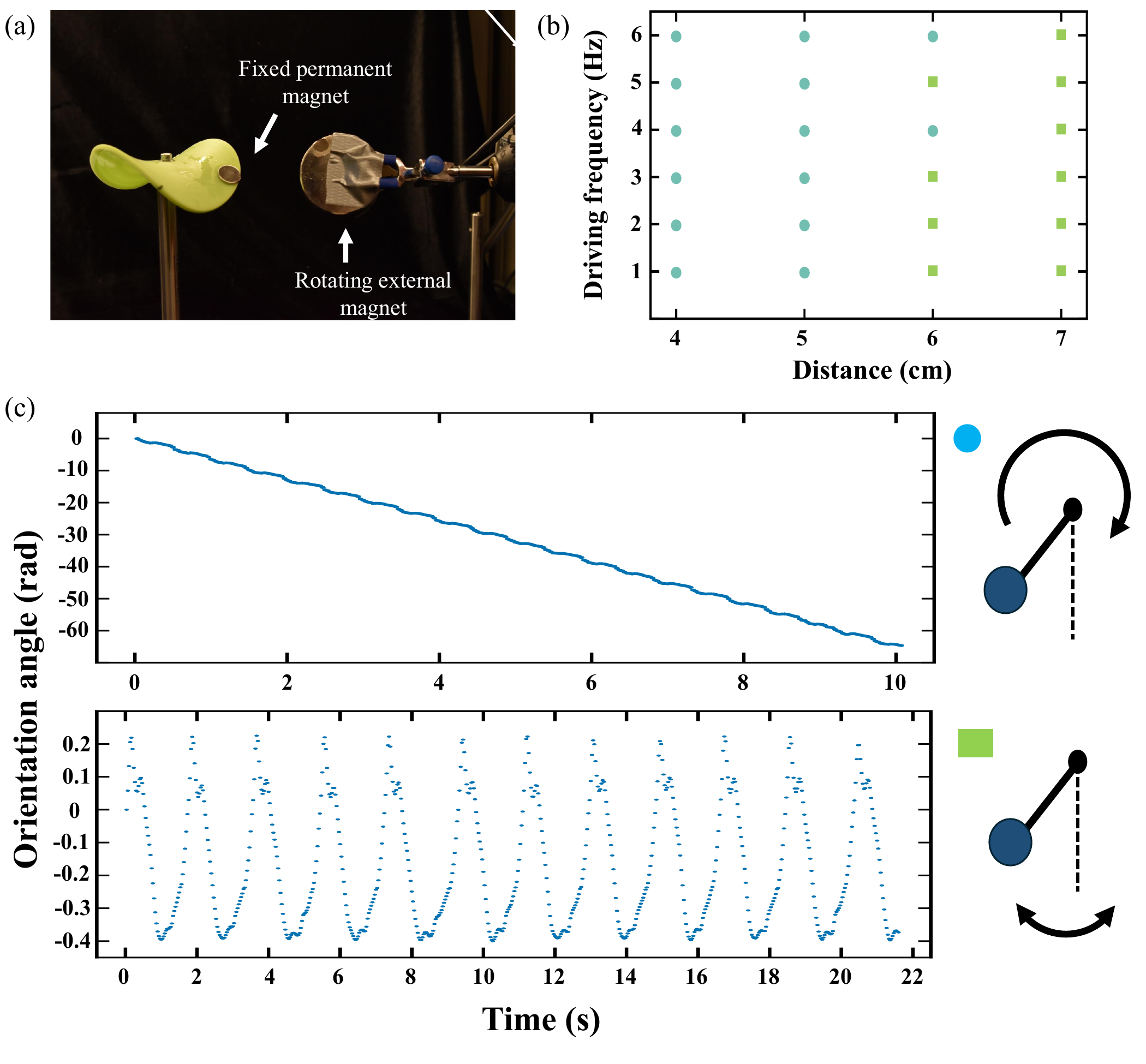}
    \caption{(a) Experimental setup for the magnetic driving. (b) Modes observed as a function of the rotating magnet's frequency $f$ and initial distance from the edge of the sheet $d$. Blue disks: spinning mode. Green squares: bounded-oscillation mode. (c) Experimental traces of the orientation angle. Clockwise spinning mode with $f = 4$ Hz and $d = 5$ cm. Bounded oscillations with $f = 1$ Hz and $d = 7$ cm.}
    
    \label{fig 2}
\end{figure}

The observed dynamics depended on the initial distance and the driving frequency (Fig.~\ref{fig 2}). For small $d$ (4 - 5 cm), the principal curvature directions transitioned between steady rotation with a regular rate to interruptions by an irregular wobbling as the driving frequency increased (see SI Sec~1.4 and SI Movies 4-7). These features appeared as large spinning-like modes in the measured orientation angle. For larger $d$ (6 - 7 cm), however, the orientation angle underwent bounded oscillations. Assuming a dipole interaction between the magnet on the surface and the rotating magnet, numerical solutions of Eq.~(\ref{equation of motion pendulum}) qualitatively reproduce this behavior using realistic experimental parameters (see SI Sec.~4.2).

\begin{figure}[h!]
    \centering
    \includegraphics[width=1\linewidth]{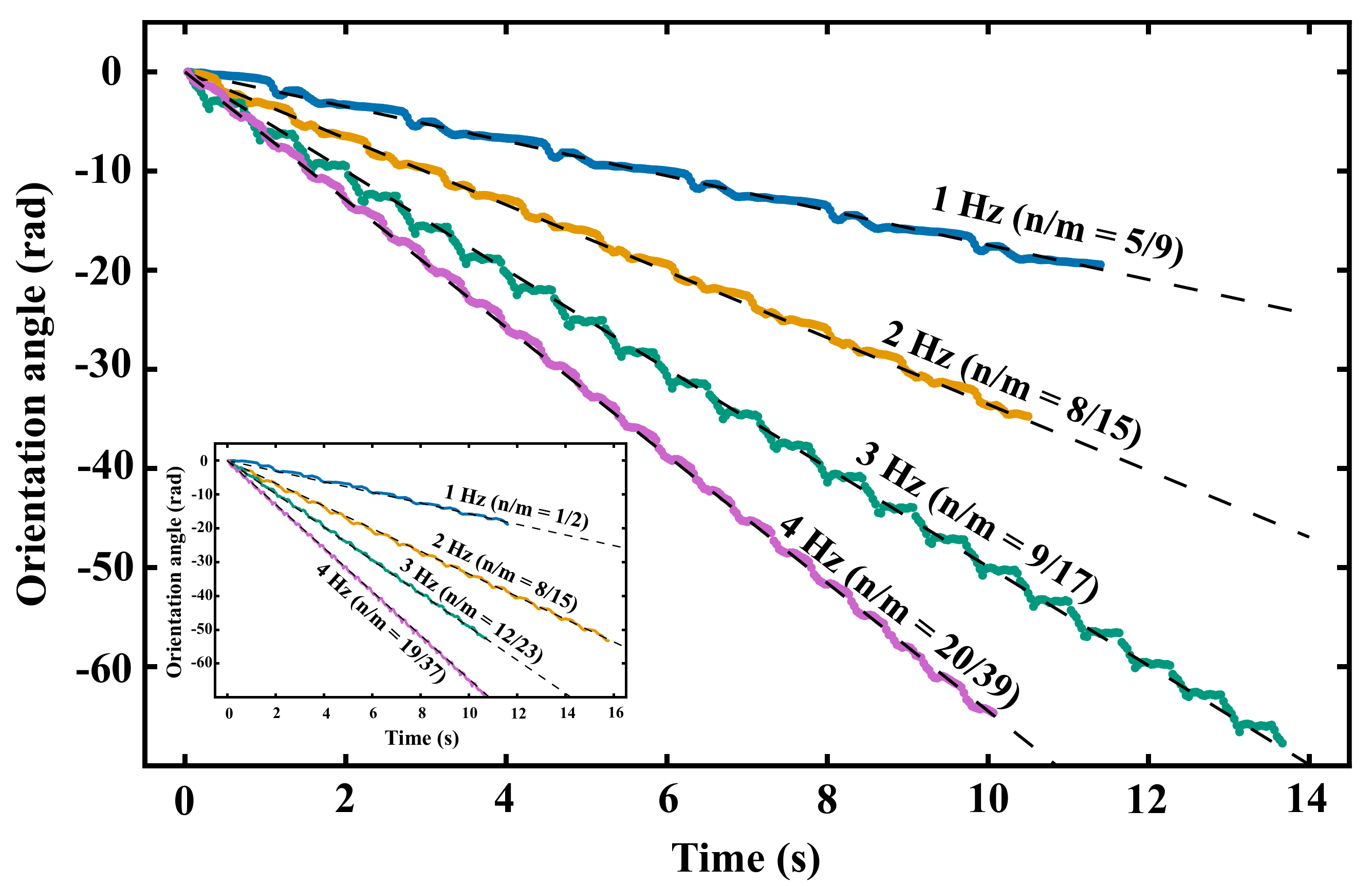}
    \caption{Spinning modes at fixed distance ($5$ cm) for different driving frequencies. Dashed lines are given by $\theta(t) = 2\pi (f/2) (n/m) t$,  showing rational phase-locking with $n/m \approx 1/2$ consistent with parametric resonance in both aged and fresh (\emph{inset}) sheets.}
    \label{fig:4}
\end{figure}

The experimental data showed a boundary at $d \approx 5-6$ cm in the $d-f$ parameter space (Fig.~\ref{fig 2}b), separating spinning modes from modes with bounded oscillations. The sheet's aging and its magnet's placement also controlled the transition distance in experiments (see SI  Sec. 1). The spinning modes exhibited rational phase-locking, where the system goes through one cycle for every $n/m$ cycles of the rotating magnet with $n, m \in \mathbb{Z}$. Most of our experiments were well described by $\theta(t) = 2\pi (f/2) (n/m) t$ with $n/m \approx 1/2$, suggesting a parametric resonance in the dynamics of the Bonnet angle for both fresh and aged sheets (Fig.~\ref{fig:4}).

To understand the origin of these features, we perform a small-angle analysis of the pendulum model with the magnetic driving, leading to a dimensionless forced Mathieu equation for $\theta \approx \delta \theta$ (see SI Sec.~4.3)
\begin{equation}
\label{Mathieu eq}
    \ddot{\delta\theta} + 2\epsilon \mu \dot{\delta\theta} + \left[1 + \epsilon \gamma\cos \left(\frac{\omega}{\omega_0}\bar{t}\right)\right]\delta \theta = \epsilon F \cos \left(\frac{\omega}{\omega_0} \bar{t}\right),
\end{equation}
written in terms of the dimensionless time $\bar{t} = \omega_0 t$. We identify the bounded oscillatory and unbounded exponentially-growing solutions of Eq.~(\ref{Mathieu eq}) with experimental oscillatory and spinning modes, respectively. The driving coefficients $\gamma$ and $F$ arise from the magnetic interaction, and hence encapsulate the dependence on distance and magnet's placement. Their explicit expressions can be found in the SI Sec.~4.3. In our experiments, the dimensionless number is approximately $\epsilon = r_m/\mathrm{R} \approx 0.16$, setting Eq.~(\ref{Mathieu eq}) in the limit of weak damping and forcing.

This limit was studied in detail in Refs. \cite{ramakrishnan2012resonances, ramakrishnan2022primary}, where it was shown that the subharmonic resonance $f \approx 2 f_0$ produces a large mechanical response in the system through a parametric resonance, in agreement with experiments. The linear modes are unbounded inside an unstable region in parameter space bounded by the curves $f = 2 f_0 + \epsilon \sigma_\pm /(2\pi)$, where $\sigma_\pm = \pm\sqrt{\gamma^2/16 - \mu^2}$. Spinning modes are confined to this region, while oscillatory modes are in its vicinity. Fig.~\ref{fig 2}b is consistent with this region with a prescribed Bonnet angle $\bar{\theta} \approx 1.37$ (see SI Sec.~4.3). Lastly, the unstable region is sensitive to the sheet's aging, becoming arbitrarily large as the sheet approaches the non-Euclidean limit, explaining the behavior of the transition distance observed in experiments.

An additional deformation mode emerged at high applied frequencies, in which the principal axes of curvature remained fixed, while the lobes underwent periodic modulation (SI Movie 8). This ``flapping'' mode requires additional degrees of freedom not considered here and can couple to the Bonnet isometry, inducing spinning modes (SI Movie 9). Its resemblance to a standing wave suggests employing a normal-mode analysis of the Enneper minimal surface, which we leave to future work.

In summary, we have studied dynamical zero-energy modes of geometric origin in prestressed sheets. By examining the elastodynamics of Enneper non-Euclidean plates, we have shown that prestresses are necessary to induce a material-independent continuous symmetry in the elastic energy of thin sheets that induces continuous stability. Remnants of this continuous symmetry can be observed in aged sheets in the form of a soft mode exhibiting complex dynamics, including irregular oscillations, steady spinning, and hybridization with other excitations. These features are consistent with resonance phenomena expected in systems with nonlinear oscillations \cite{nayfeh2024nonlinear}. Reasoning in the opposite direction suggests that other soft deformations in stress-free sheets can become zero modes by finding an appropriate distribution of prestresses. These results encourage examining other dynamical thin elastic systems through a geometric lens, with an eye towards designing continuum sheets that deform in prescribed ways.


We are indebted to Panagiotis Koutsogiannakis, Massimo Ruzzene, Lauren Altman, and James Hanna for insightful comments. We acknowledge funding through NSF CMMI2247095.

Experimental orientation angle trajectories, experimental relaxation times, MATLAB code for extracting orientation angles and Mathematica code for data analysis and numerical solutions can be found in \cite{Note1}. SI Movies can be found in \cite{Note2}. Additional experimental and mathematical details are found in the SI.

\subsection{Materials and Methods}

\subsection*{Materials}
A polyvinylsiloxane (PVS) resin (Zhermack Elite Double 32), was used to fabricate the sheets. Silicone oil with a viscosity of 40 cSt was obtained from Thermo Scientific. All chemicals were used as received. Nanoindentation measurements were performed on an FT-MTA03 micromechanical testing system (FemtoTools) using a spherical glass tip with a 25 µm radius. Molds were printed on a Form 3+ 3D printer using Grey Pro resin or Clear Resin V4.1 (Formlabs). 3D scans were taken using a Revopoint MINI 2 3D scanner (Revopoint 3D) and curvatures were calculated with a user-developed MATLAB code after first smoothing the mesh to reduce noise. Videos were captured with a FASTCAM SA3 120K-M2 high-speed camera (Photron). Point and angle tracking was done using a custom-built MATLAB code. The rotating magnetic field was generated by clamping a neodymium magnet (approx. 4 cm diameter by 2.5 cm thick) on an overhead mechanical stirrer (Chemglass Life Sciences) oriented horizontally.

\subsection*{Fabrication of silicone Enneper minimal surfaces}

We prepared Enneper minimal surfaces with two lobes, characterized by the areal shrinking $\Omega(r)= \left[c\left(1+(r⁄R)^{2}\right)\right]^2$, where $R$ was the characteristic radius of curvature, and $c$ was a constant fixing the value of $\Omega$ at $r = 0$. We had control over the magnitude of curvature and the number of resulting lobes (see SI Sec. 1), but chose $c = 0.7$ and $R = 12$ cm to obtain a gently curved surface for our experiments.

\begin{figure}[h!]
    \centering
    \includegraphics[width=1\linewidth]{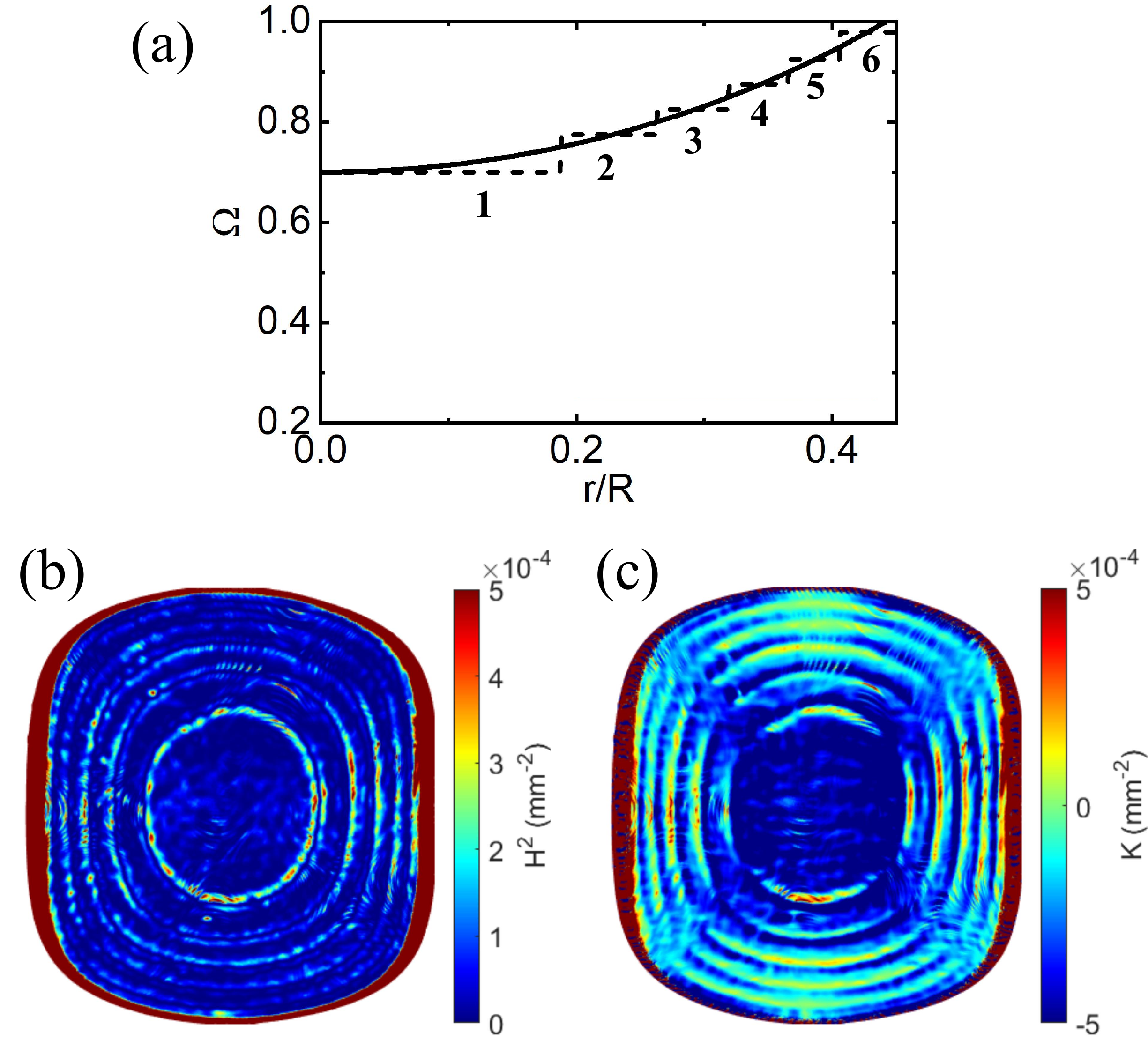}
    \caption{a) Areal shrinking and its stepwise approximation with 6 steps for the two-lobed Enneper minimal surface with $c = 0.7$ and $R = 12$ cm as a function of the ratio of the radial distance on the flat disk to the radius of curvature of the target shape, $r/R$. b)  Squared mean curvature, $H^2$, of the sample. c) Gaussian curvature, $K$, of the sample.}
    \label{fig:5}
\end{figure}

The target metric was divided into 6 steps of equal areal shrinking, $\Omega$, and the corresponding radii of the annuli were determined such that the smooth metric intersected each step in $\Omega$ at its midpoint (Fig.~\ref{fig:5}a). Molds were 3D-printed, measuring 3 mm deep and with radii corresponding to the dimensions of the desired ring. A 2 mm diameter hole was added as a mounting point at the center. The PVS was mixed with the required amount of silicone oil to achieve the value of $\Omega$ for a given step, then poured into the mold. The amount of required silicone oil needed for a given areal shrinking ratio was determined by noting that, for isotropic shrinkage, \(\Omega = \phi_0^{2/3}\) where $\phi_0$ is the PVS volume fraction, assuming complete extraction of the added silicone oil. 

Each annulus was allowed to react until it was strong enough to be removed from the mold, then placed in the next mold. All molds were printed with the same constant outer radius to accommodate the previous layers, but with a smaller inner radius, allowing the next layer to be poured. Once all 6 layers were finished, the entire sample was left overnight to ensure complete reaction. The sample was then removed from the final mold and soaked in chloroform or acetone for $\approx3$ days, replacing the solvent every $\approx1$ day. We found that chloroform, acetone, and hexane yielded the same extraction results. Then the sample was carefully taken out of the solvent bath, as it was very fragile when swollen, and allowed to air dry for 2 – 4 h. It was then dried in vacuo at room temperature for another 2 h to ensure complete solvent removal.

The resulting surface (Fig.~\ref{fig 1}b) had the expected zero mean curvature, except at the interface between rings of differing oil content, where differences in thickness manifested as a ridge of high mean and positive Gaussian curvature (Fig.~\ref{fig:5}b-c). 

\subsection*{Experimental dynamics of the Enneper minimal surface}
The sample was mounted on an optical post with a screw, and a camera was positioned directly overhead, pointing down at the top surface of the sample. 

To study the free dynamics of the sheet, the sample was manually manipulated such that the principal curvature directions were turned $\pi/2$ from their preferred orientation, then released and allowed to relax back, while video was being captured. The orientation angle in each frame was determined by binarizing the image to isolate the sample, then calculating the orientation of the major axis of an ellipse with the same second moment of mass distribution as the sample (accomplished by a custom-built MATLAB code using the built-in “regionprops” function). Point tracking was done with a custom-built MATLAB code. 

To study the forced dynamics of the sheet, a cylindrical neodymium magnet (approx. 1.5 cm diameter by 1 mm thick) was superglued to a point on the edge of the sheet, while a larger cylindrical neodymium magnet ($\approx4$ cm diameter by 2.5 cm thick) was clamped on the overhead stirrer using a custom 3D printed holder and positioned $\approx 4 - 7$ cm away from the edge of the Enneper minimal surface, and at the same height as the surface's center. The motor was run at frequencies ranging from 1 – 6 Hz, and videos were captured. Orientation angles were measured in the same way as before with during the free dynamics. 

See SI Sec. 1 for measurements of the orientation angle during forced dynamics and representative orientation angles during free dynamics.

\subsection*{Estimating the natural frequency and relaxation time of the sample from the free dynamics}

In the underdamped limit, the solution to Eq.~(\ref{equation of motion pendulum}) when $\theta \approx \delta \theta$ is small is
\begin{equation}
    \delta \theta = A e^{-\frac{t}{t_\text{relax}}}\cos\left(\omega_D t + \varphi_0\right),
\end{equation}
where $A$ and $\varphi_0$ are constant amplitude and possible phase shift, $\omega_0^2 = h^3\bar{k}/\mathrm{I}$ is the characteristic angular velocity of the system, and $\omega_D = \sqrt{\omega_0^2 - 1/t_\text{relax}^2}$ is the observed angular velocity of the oscillations. The observed and natural frequencies from the data are $f_D = \omega_D/(2\pi)$ and $f_0 = \omega_0/(2\pi)$, respectively.

We extracted $t_\text{relax}$ and $f_0$ from the experimental orientation angles by restricting the data to the oscillatory regime of motion as the angle approached its equilibrium value, and by finding the times $T_n$ at which the n-th peak occurred. The peaks occurred at $T_n = 2\pi n/ (\omega_D) = n/f_D$, where $n$ is an integer, and laid on the exponential envelope of the solution $\delta\theta(T_n) = A e^{-T_n/t_\text{relax}}$. We extracted $t_\text{relax}$ from an exponential fit to these data points. To estimate the natural frequency, we first estimated $f_D$ as $f_D \approx N/ \Delta T $, where $\Delta T = T_{\text{last}} - T_{\text{initial}}$ is the difference between the times of the last and initial peak and $N$ is the number of peaks in between. From this and the extracted $t_{\text{relax}}$, the definition of $f_D$ gave an estimate of $f_0$. See SI Sec. 3.4 for a histogram of the estimated values.

\subsection*{Bonnet isometry of the Enneper minimal surface}
The associate family of a minimal surface can be written in the form $\mathbf{X} = \cos \theta ~ \mathbf{X}_0 + \sin \theta ~ \mathbf{X}_C$, where $\theta$ is the Bonnet angle parameterizing the Bonnet isometry. Employing a system of polar coordinates, $\left(r, \phi\right)$, around the center of the plate, the Bonnet isometry of Enneper minimal surface can be parameterized as \cite{dierkes2010minimal}
\begin{align}
x(r, \phi; \theta) &= c\left(r\cos\left( \theta + \phi \right) - \frac{r^3}{3 R^2}\cos\left(\theta + 3\phi\right)\right),\\
y(r, \phi; \theta) &= -c\left(r\sin\left( \theta + \phi \right) + \frac{r^3}{3 R^2}\sin\left(\theta + 3\phi\right)\right), \\
z(r, \phi; \theta) &= \frac{c r^2}{R}\cos\left(\theta + 2\phi\right),
\end{align}
where $R$ is the characteristic radius of curvature of the shell and $c$ is a proportionality constant. We have suppressed the time dependence of the angle $\theta(t)$ for brevity. 

The metric tensor of the Enneper minimal surface in this system of coordinates takes the form
\begin{equation}
g_{ij}\mathrm{dx}^i\mathrm{dx}^j = c^2\left(\frac{r^2}{\mathrm{R}^2} + 1\right)^2\left(\mathrm{dr}^2 + r^2 \mathrm{d\phi}^2\right),
\end{equation}
where we sum over repeated indices that range over the coordinates $\left(r, \phi\right)$.

It is straightforward to compose Bonnet isometries. Starting with $\mathbf{X}\left(\bar{\theta}\right) = \cos \bar{\theta} ~ \mathbf{X}_0 + \sin \bar{\theta} ~ \mathbf{X}_C$, its conjugate surface is obtained by mapping $\bar{\theta} \mapsto \bar{\theta} + \pi/2$. Thus, $\mathbf{X}_C\left(\bar{\theta}\right) = -\sin \bar{\theta} ~ \mathbf{X}_0 + \cos \bar{\theta} ~ \mathbf{X}_C$. We can apply a second Bonnet isometry with angle $\theta$ to write
\begin{eqnarray}
    \mathbf{X}(\theta, \bar{\theta}) &=& \cos \theta ~ \mathbf{X}\left(\bar{\theta}\right) + \sin \theta ~ \mathbf{X}_C\left(\bar{\theta}\right), \\
    &=& \cos \left(\theta + \bar{\theta}\right) ~ \mathbf{X}_0 + \sin \left(\theta + \bar{\theta}\right) ~ \mathbf{X}_C,\\
    &=& \mathbf{X}(\theta + \bar{\theta}).
\end{eqnarray}
Hence, the composition of Bonnet isometries is additive in Bonnet angles. The prescribed Bonnet angle is $\bar{\theta}$ while $\theta$ parameterizes the deformation by a Bonnet isometry away from $\bar{\theta}$.

\subsection*{Removing Rigid-body motions from the Bonnet isometry}
Our experimental setup fixed the plate's center, excluding rigid-body motions. The Bonnet isometry, however, does not account for this. Note that the radial tangent vector at the origin during the deformation is given by
\begin{equation}
    \partial_r \mathbf{X} = c \left( \cos\left[\theta + \bar{\theta} + \phi\right] \bold{\hat{x}}  -\sin\left[\theta + \bar{\theta}  + \phi\right] \bold{\hat{y}}\right),
\end{equation}
which changes over time when $\theta(t)$ is dynamical. To exclude rigid-body motions, we fix $\partial_r\mathbf {X}$ for all time. We compose the Bonnet isometry with a rotation around the $z$ axis by an angle $\alpha(\theta)$ that depends on $\theta$, $\mathrm{R}_\alpha \left[\mathbf{\hat{z}}\right]$ 
\begin{equation}
\mathbf{X}(r, \phi; \theta) = \mathrm{R}_\alpha \left[\mathbf{\hat{z}}\right]\left(\cos \left[\theta + \bar{\theta}\right] \mathbf{X}_0 + \sin \left[\theta + \bar{\theta}\right] \mathbf{X}_C\right).
\end{equation}
The radial tangent vector at $r = 0$ now satisfies
\begin{align}
\partial_t\left(\partial_r\mathbf{X}\right) =  c\left(\alpha'(\theta) - 1\right)\partial_t \theta\left(\sin\left[\theta + \bar{\theta} - \alpha\right]\bold{\hat{x}} + \right. \nonumber \\
\left. \cos\left[\theta + \bar{\theta} - \alpha\right] \bold{\hat{y}}\right),
\end{align}
where $\partial_t$ denotes time derivative. Choosing $\alpha = \theta$ guarantees the radial tangent at the origin remains constant over time.

Explicit computations show that the second fundamental form, $b_{ij}$, and metric, $g_{ij}$, of the Enneper minimal surface do not change under the rotation. However, the original unit normal vector, $\hat{\mathbf{N}}_{\bar{\theta}}$, becomes $ \hat{\mathbf{N}} 
= \mathbf{R}_\theta \hat{\mathbf{N}}_{\bar{\theta}}$ and it is explicitly given by
\begin{equation}
\hat{\mathbf{N}} = \frac{2rR\cos(\theta + \phi)}{r^2 + R^2} \bold{\hat{x}} +  \frac{2rR\sin(\theta + \phi)}{r^2 + R^2} \bold{\hat{y}} + \frac{r^4 - R^4}{\left(r^2 + R^2\right)^2} \bold{\hat{z}}.
\end{equation}

\bibliography{references}

\end{document}


\title[Article Title]{Supplementary Information - Dynamical geometric modes in non-Euclidean plates}


\author[1]{\fnm{Joseph C.} \sur{Roback}}\email{joseph.roback@colorado.edu}

\author[2]{\fnm{Carlos E.} \sur{Moguel-Lehmer}}\email{cmoguell@syr.edu}

\author[3]{\fnm{Katharina A.}\sur{Fransen}}\email{Katharina.Fransen@colorado.edu}

\author*[2]{\fnm{Christian D.} \sur{Santangelo}}\email{cdsantan@syr.edu}

\author*[1,3]{\fnm{Ryan C.} \sur{Hayward}}\email{ryan.hayward@colorado.edu}

\affil[1]{\orgdiv{Department of Chemical and Biological Engineering}, \orgname{University of Colorado Boulder}, \orgaddress{\city{Boulder}, \postcode{80303}, \state{Colorado}, \country{USA}}}

\affil[2]{\orgdiv{Department of Physics}, \orgname{Syracuse University}, \orgaddress{\city{Syracuse}, \postcode{13244}, \state{New York}, \country{USA}}}

\affil[3]{\orgdiv{Materials Science and Engineering Program}, \orgname{University of Colorado Boulder}, \orgaddress{\city{Boulder}, \postcode{80303}, \state{Colorado}, \country{USA}}}

\maketitle

\tableofcontents

\pagebreak 

\section{Experimental Methods}\label{sec1}

\subsection{Characterization of silicone oil extraction}

To trigger a change in volume, the polyvinylsiloxane resin was diluted with silicone oil, which does not participate in the cross-linking reaction, prior to curing. After curing, the material was soaked in a solvent to remove the silicone oil, resulting in volumetric shrinkage. By varying the amount of added silicone oil, the magnitude of the shrinkage could be controlled. To characterize the amount of extractable material, we fabricated several small discs with PVS volume fractions, $\phi_0$, ranging from 0.2 to 1. It was found that, for values of $\phi_0$ below approximately 0.2, robust gels that could be reliably handled and removed from the molds were not formed, setting an approximate lower limit on achievable shrinking magnitude. Molds of discs measuring 10 mm in diameter and 3 mm deep were 3D printed and filled with the appropriate composition of PVS and silicone oil. The discs were allowed to cure overnight, then they were weighed, soaked in chloroform for 24 h to extract the added silicone oil, allowed to fully dry in air, then finally weighed again. We found that acetone and hexane as washing solvents also gave the same results as chloroform. The ratio of washed mass, $m$, to unwashed mass, $m_0$, was calculated (Fig.~\ref{Fig: S1}), where the black line traces \(y=x\), representing extraction of exactly the amount of added silicone oil. These data indicate that approximately all the added silicone oil was extracted, and the gel fraction of the pure green silicone was nearly one.

\makeatletter 
\renewcommand{\thefigure}{S\@arabic\c@figure}
\makeatother

\begin{figure}[h!]
\centering
\includegraphics{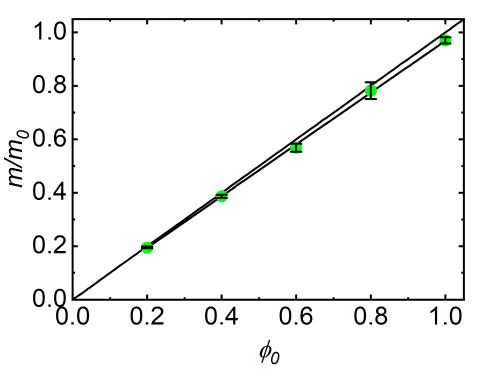}
\caption{Ratio of mass of green silicone discs before and after washing as a function of $\phi_0$.}
\label{Fig: S1}
\end{figure}

\pagebreak

\subsection{Fabrication of silicone Enneper minimal surfaces}

By varying the amount of added silicone oil in an annulus-by-annulus approach, we could control the curvature and number of lobes of the final surface. 
To exemplify this, we prepared Enneper’s surfaces with two or three nodes, which were given by the areal shrinking \(\Omega(r)=[c\left(1+(r⁄R)^{2(n-1)} \right)]^2\), where $n$ was the number of nodes, $R$ was a characteristic radius of curvature, and $c$ was a constant that sets the value of $\Omega$ at $r = 0$. We initially chose $c = 0.34$, which corresponds to the highest magnitude of volumetric shrinkage accessible ($m/m_0 \approx 0.2$), and $R = 60\ \mathrm{mm}$, which allowed for a significant portion of the surface to be generated using an initial disc radius of $50 – 60 \mathrm{mm}$.

\begin{figure}[H]
\centering
\includegraphics{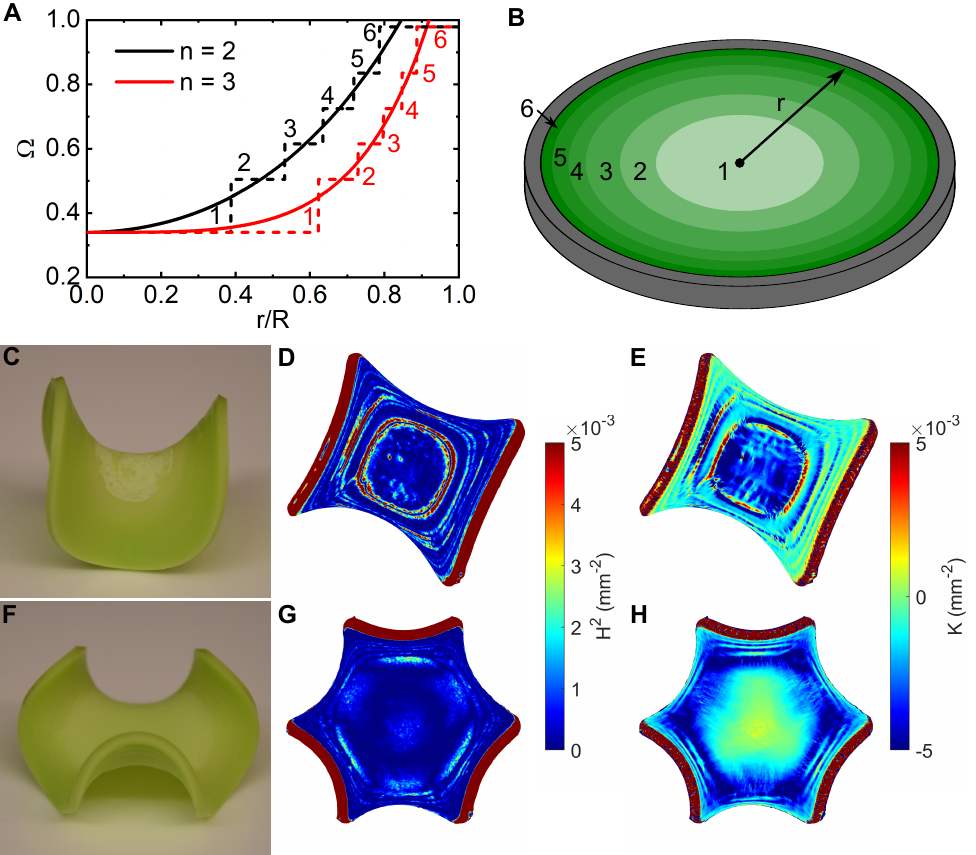}
\caption{A) areal shrinking and stepwise approximation for two- and three-node Enneper’s surfaces with $c$ = 0.34 and $R$ = 60 B) Schematic for fabrication of silicone Enneper’s surfaces. C), D), and E) Photograph, squared mean curvature, $H^2$, and Gaussian curvature, $K$, respectively, of two-node Enneper’s surface. F), G), and H) Photograph, squared mean curvature and Gaussian curvature, respectively, of three-node Enneper’s surface.}
\label{S2}
\end{figure}

\subsection{Free dynamics of Enneper minimal surface}

\begin{figure}[H]
    \centering
    \includegraphics[width=1\linewidth]{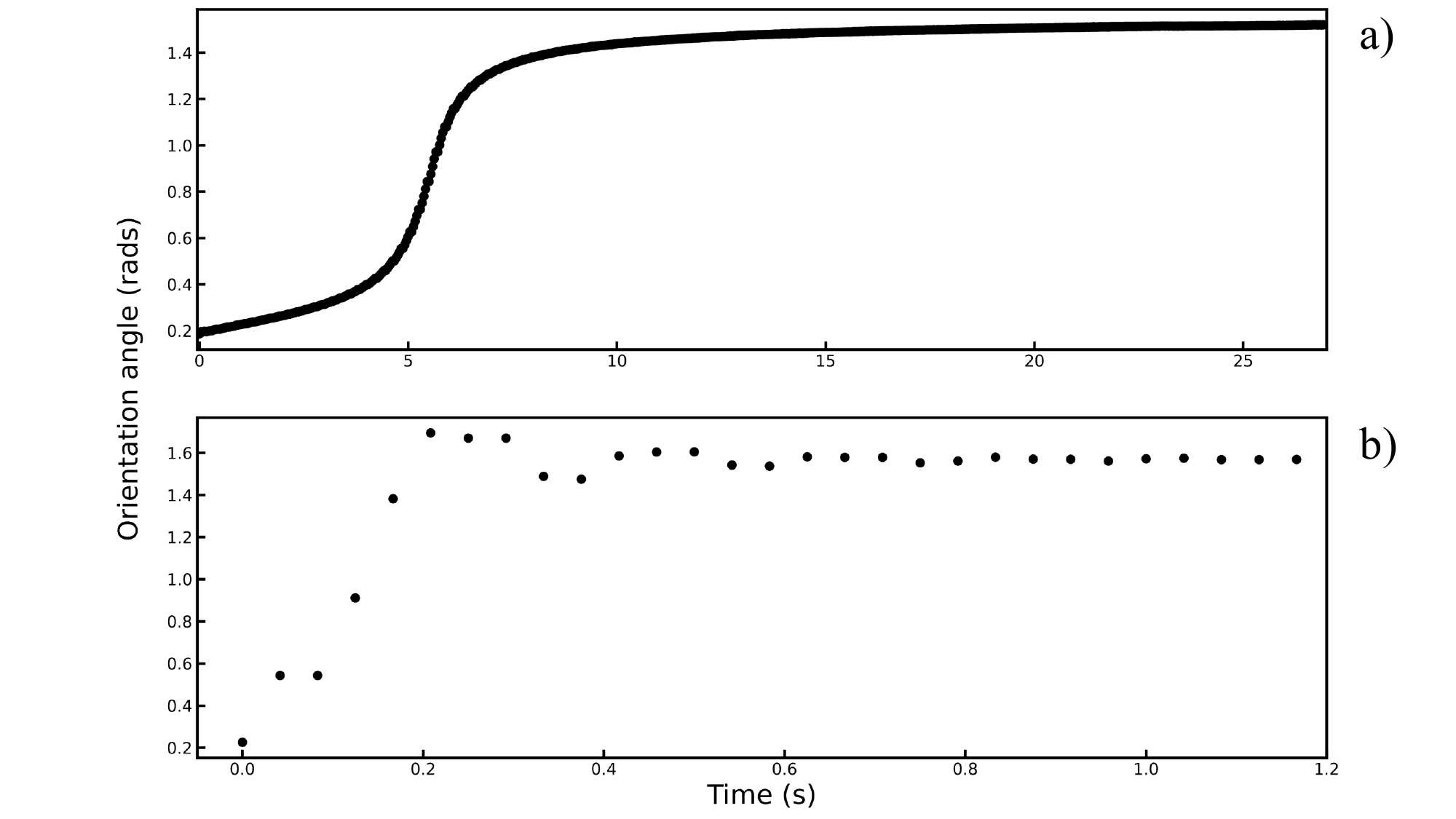}
    \caption{Orientation angle after deforming the sheet's lobes away from their equilibrium orientation, which is at $\pi/2$. a) The sheet shows overdamped relaxation 24 hours after fabrication. b) The sheet shows underdamped relaxation 120 hours after fabrication.}
    \label{fig:S4}
\end{figure}

Fig.~\ref{fig:S4} shows representative measurements of the orientation angle.  See \cite{Note1} for the remaining measured orientation angles and MATLAB code.

\subsection{Magnetic driving of Enneper minimal surface}

\begin{figure}[H]
    \centering
    \includegraphics[width=1\linewidth]{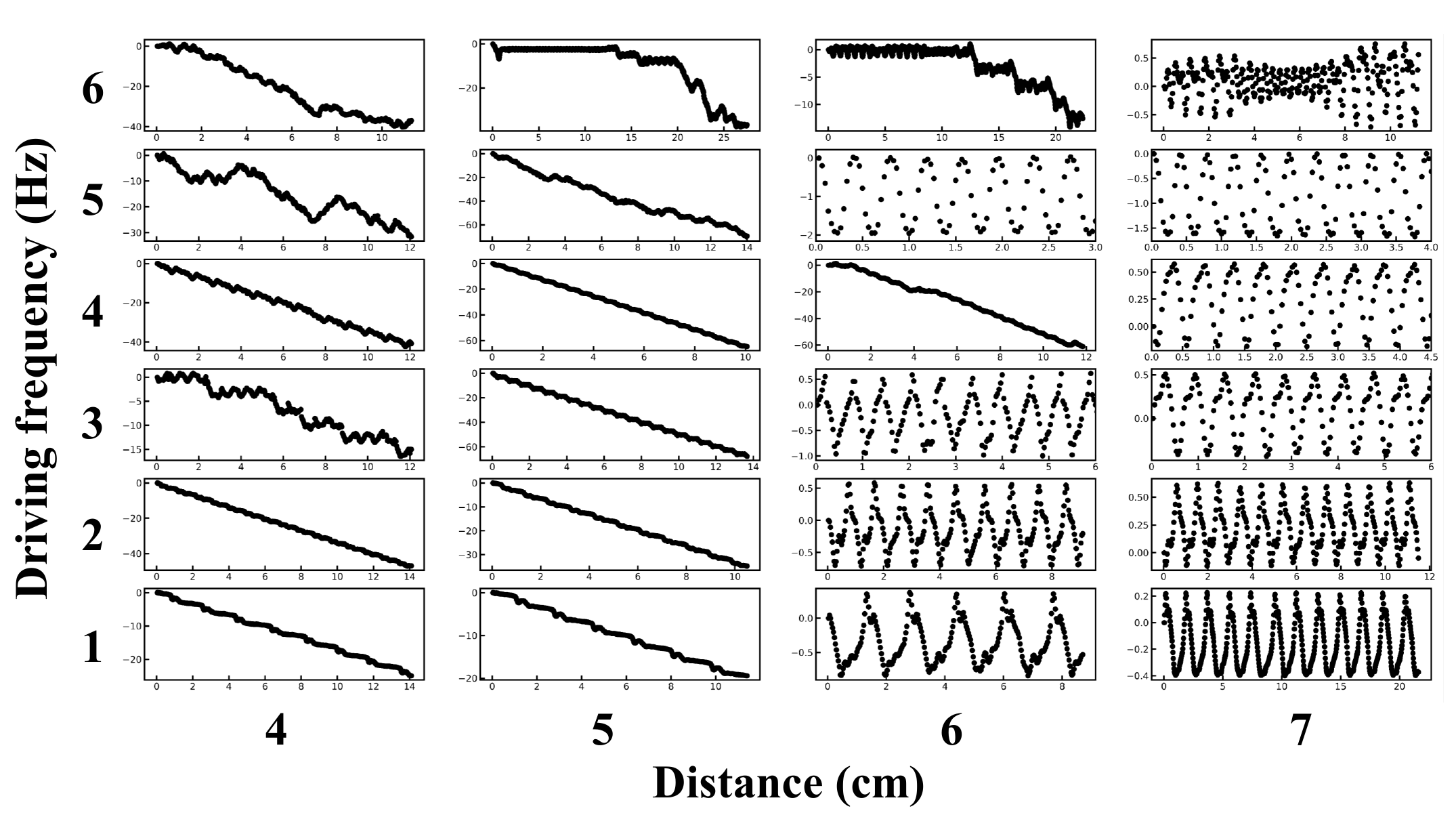}
    \caption{Measured orientation angles (rads) over time (s) for different driving frequencies and distances, with the sheet's magnet at $\phi_0 \approx 0$ and $r_0 \approx \mathrm{R}$ on the surface, showing a clear transition from spinning modes to bounded oscillations at $d \approx 5-6$ cm.  Spinning occurs clockwise. This dataset corresponds to the phase diagram in the main text.}
    \label{fig:Magnetic driving data}
\end{figure}

\begin{figure}[H]
    \centering
    \includegraphics[width=1\linewidth]{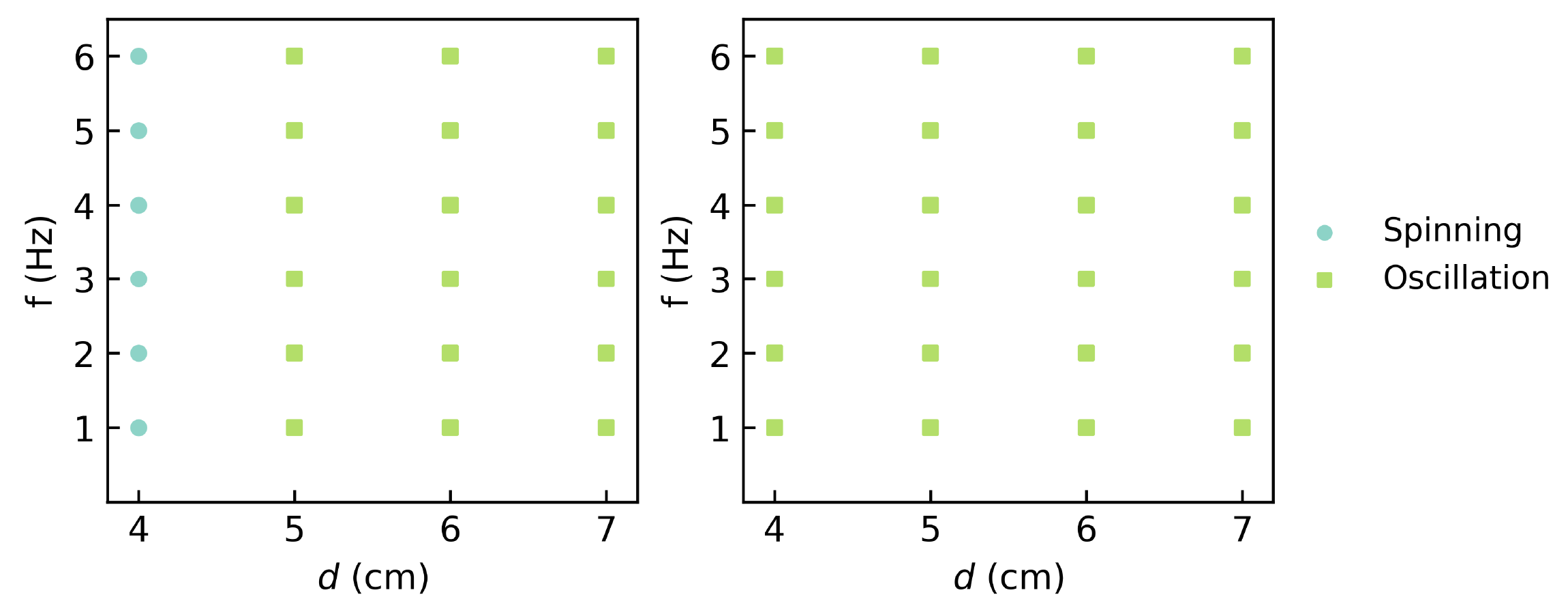}
    \caption{Phase diagrams of pendulum modes as a function of the magnet's position on the sheet, $\phi_0$. We keep the magnet close to the edge of the sheet, $r_0 \approx \mathrm{R}$. Left: $\phi_0 \approx \pi/4$. Right: $\phi_0 \approx \pi/2$. }
    \label{fig:angle dependence}
\end{figure}

\begin{figure}[H]
    \centering
    \includegraphics[width=0.6\linewidth]{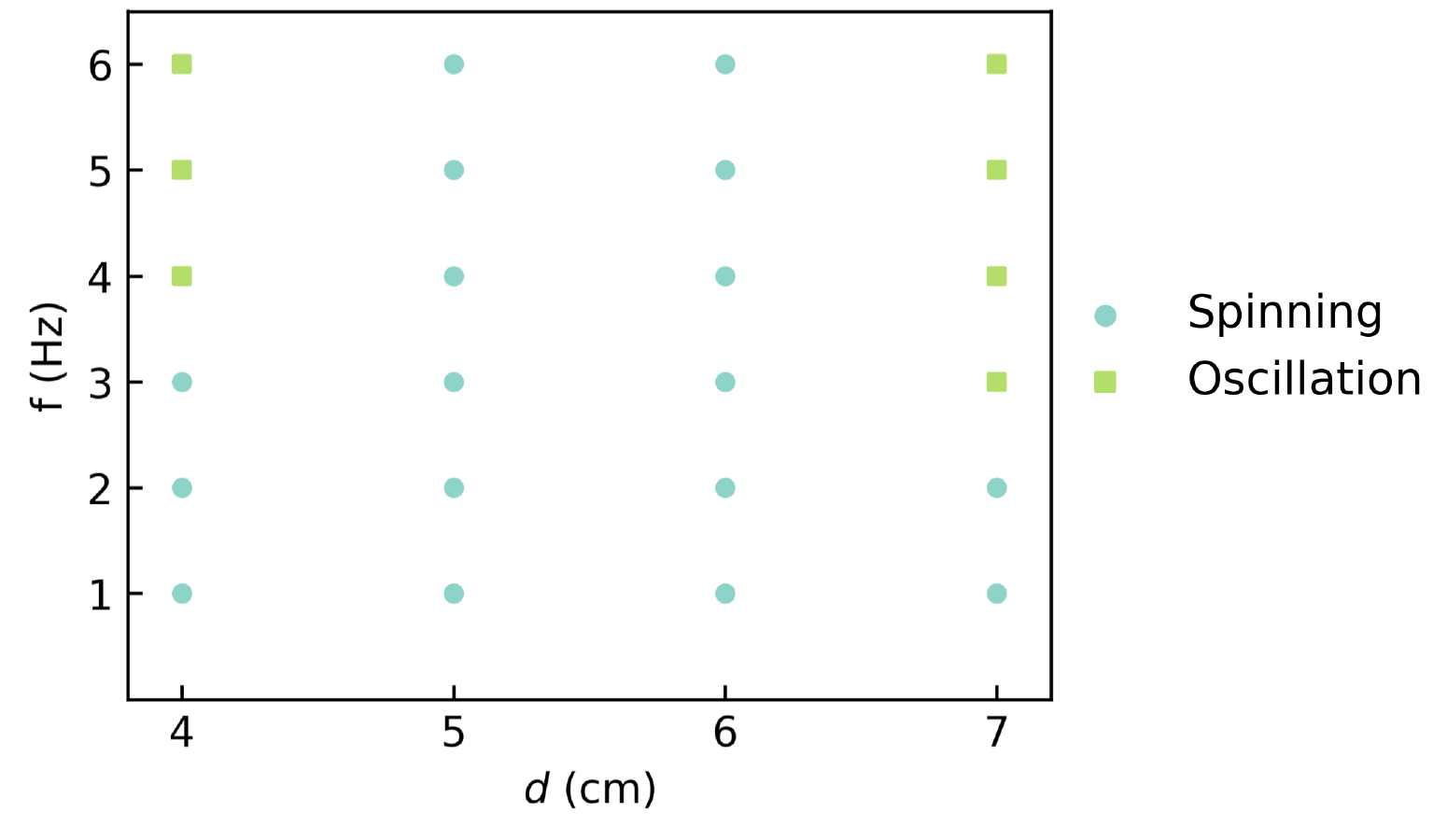}
    \caption{Phase diagram of the pendulum modes a few hours after the sheet was fabricated. The sheet's magnet position is $\phi_0 \approx 0$ and $r_0 \approx \mathrm{R}$. The oscillating modes are flapping modes.}
    \label{fig:aging dependence}
\end{figure}

Orientation angles are shown in Fig.~\ref{fig:Magnetic driving data}. Fig.~\ref{fig:angle dependence} shows how the spinning-oscillations phase diagram changed as a function of the position of the magnet on the sheet, while Fig.~\ref{fig:aging dependence} shows how phase diagram looked for a freshly-made sheet. See \cite{Note1} for measured orientation angles.

\pagebreak 

\section{Deformations of Enneper's minimal surface}

\subsection{Removing rigid-body rotations}

\begin{figure}[H]
    \centering
    \includegraphics[width=1\linewidth]{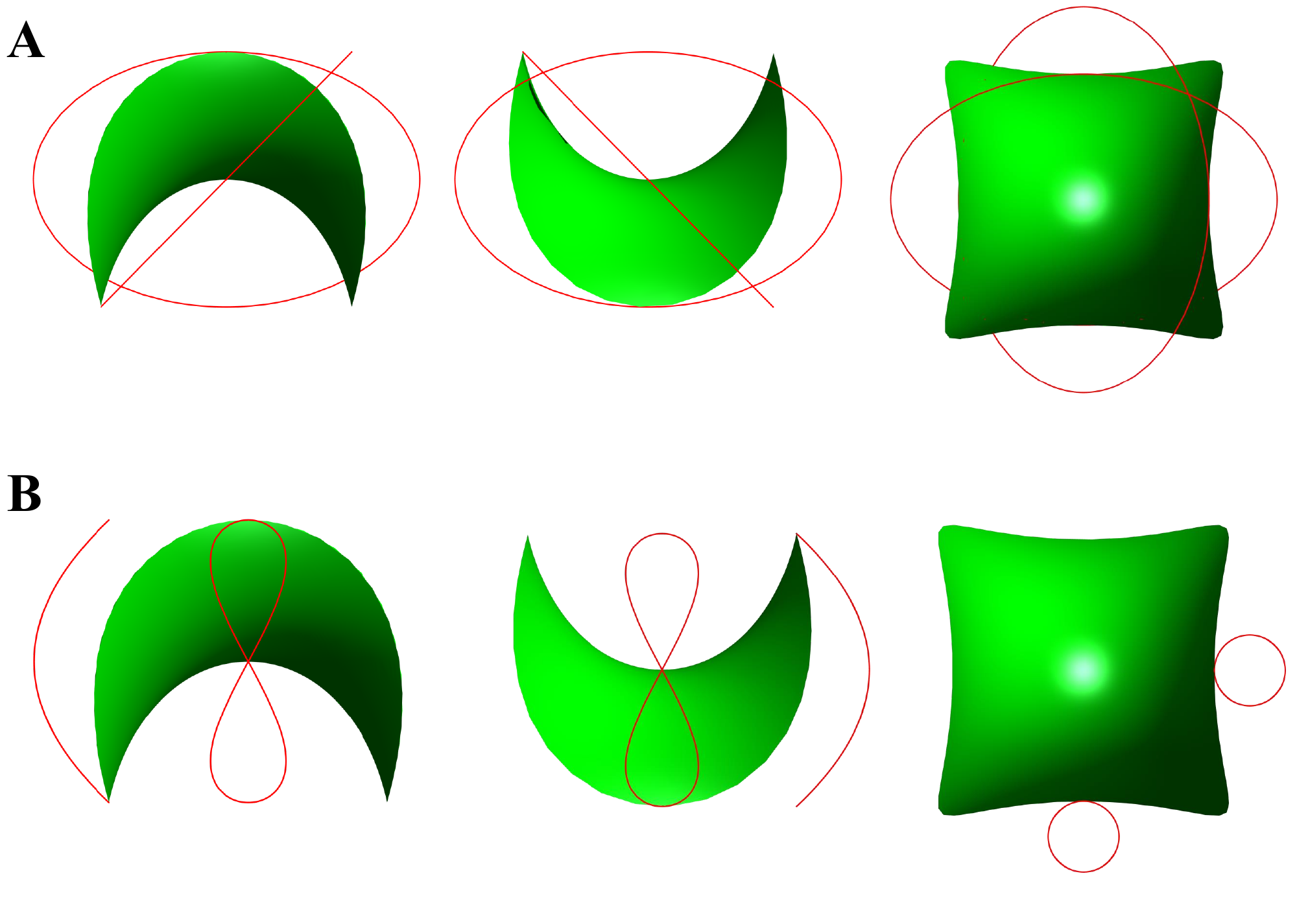}
    \caption{Example of material point trajectories under the Bonnet isometry on the Enneper minimal surface in red. A) Trajectories follow elliptical paths before applying the counter-rotation $\mathrm{R}_\theta \left[\mathbf{\hat{z}}\right]$. B) Trajectories follow eight-figure paths after removing rigid-body motions with the counter-rotation.}
    \label{fig:S3}
\end{figure}

The experimental setup excluded rigid-body motions by fixing the plate's center. The Bonnet isometry does not account for this, and the material points on the surface describe elliptical orbits. Composing the Bonnet isometry with a rotation $\mathrm{R}_\theta \left[\mathbf{\hat{z}}\right]$ around the $z$ axis by the angle $\theta$ removes the rigid-body rotations with material points describing figure eights instead (Fig.~\ref{fig:S3}). 

\subsection{Bending energy under the Bonnet isometry}

The bending energy for the midsurface of a thin shell can be written in terms of the bending moment, $m_{ij} = \frac{1}{2}(b_{ij} - \bar{b}_{ij})$ as \cite{efrati2009elastic}
\begin{equation}
\mathcal{E}_b = \frac{h^3}{2} \int \mathrm{dA}  \mathrm{dt} \ m_{ij}M^{ijkl}m_{kl}
\label{Bending}
\end{equation}
where $h$ is the shell thickness, $b_{ij}$ is the second fundamental form of the midsurface, $\bar{b}_{ij}$ is the prescribed second fundamental form, and $M^{ijkl} = \kappa\Bar{g}^{ij}\Bar{g}^{kl} + 2\bar{\kappa}\Bar{g}^{ik}\Bar{g}^{jl}$ is expressed in terms of the bending rigidities, $\kappa$ and $\bar{\kappa}$ of the sheet. Non-euclidean plates satisfy $\bar{b}_{ij} = 0$. 

In isothermal coordinates, the second fundamental form of any member of the associate family is given by \cite{moguel2025topological}
\begin{equation}
\label{Curvature tensor Bonnet}
b_{ij} = \cos\left[\theta + \bar{\theta}\right] b^0_{ij} + \sin\left[\theta + \bar{\theta}\right] b^C_{ij},
\end{equation}
where $b^0$ and $b^C$ are the second fundamental forms of $\mathbf{X}_0$ and $\mathbf{X}_C$, respectively. We assume that the prescribed second fundamental form takes this form as well with the prescribed Bonnet angle, $\bar{\theta}$. In the principal frame of $\mathbf{X}_0$, the second fundamental forms can be written as
\begin{equation}
{b}_{ij}^0 = k \ \Omega 
\begin{pmatrix}
    1 && 0\\
    0 && -1
\end{pmatrix}, \quad 
{b}_{ij}^C = k \ \Omega 
\begin{pmatrix}
    0 && 1\\
    1 && 0
\end{pmatrix},
\label{second forms principal frame}
\end{equation}
where $k$ is the principal curvature of the surface. Using these representations, and assuming that the induced shape is the member of the associate family at $\theta + \bar{\theta}$, the bending energy is
\begin{align}
\mathcal{E}_b &= \frac{h^3}{6} \int \mathrm{dA} \ m_{ij}M^{ijkl}m_{kl}, \nonumber\\
&= \frac{h^3}{24}\int \mathrm{dA} \left[\kappa \left(\Bar{g}^{ij}b_{ij} - \Bar{g}^{ij}\Bar{b}_{ij}\right)^2 +  2\bar{\kappa} \left(b_{ij}b^{ij} - 2b_{ij}\bar{b}^{ij} + \Bar{b}_{ij}\Bar{b}^{ij}\right)\right], \nonumber \\
&= \frac{h^3\bar{\kappa}}{3}\left[\int \mathrm{dA}\left(- \mathrm{K}\right) \right] ~ \left(1 - \cos\theta\right),
\label{Bending Constant Bonnet}
\end{align}
where $\mathrm{K}$ is the Gaussian curvature of the midsurface.

\section{Equation of motion for the Bonnet angle from a 3D viscoelastic model}

\subsection{Kinematic equations}

We write the action, $\mathcal{S}$, of the three-dimensional system in terms of an arbitrary embedding of the shape, $\boldsymbol{\chi}$
\begin{equation}
\mathcal{S} = \int \mathrm{dt}\mathrm{dV} ~ \left[\frac{\rho}{2}\left(\partial_t \boldsymbol{\chi}\right)^2 - \sigma^{ij}_{\text{Total}}\gamma_{ij}^{\text{3D}}\right],
\label{action 3D KV}
\end{equation}
The first term is the kinetic energy of the deformation with mass density $\rho$. The second term is the total elastic energy of the system with a stress tensor, $\sigma_{\text{Total}}$, and its strain tensor in three dimensions, $\gamma_{ij}^{\text{3D}} = \frac{1}{2}\left(g_{\text{3D}} - \bar{g}_{\text{3D}}\right)$,  which is written using the three-dimensional metric tensor of the shell, $g_{\text{3D}}$, and a prescribed metric tensor, $\bar{g}_{\text{3D}}$, that makes the  shell stress-free. The prescribed volume element, $\mathrm{dV} = \mathrm{dx^1}\mathrm{dx^2}\mathrm{dx^3}\sqrt{\mathrm{\det}\bar{g}}$, is written in an arbitrary coordinate system, $\left(x^1, x^2, x^3\right)$, using $\bar{g}_{\text{3D}}$. The induced metric in the strain tensor is written in terms of the arbitrary embedding as $g_{ij}^{\text{3D}} = \partial_i \boldsymbol{\chi} \cdot \partial_j \boldsymbol{\chi}$, where $\partial_i$ denotes the derivative with respect to the $x^i$ coordinate. We restrict the allowed embeddings $\boldsymbol{\chi}$ by imposing that the midsurface of the shell must conform to a surface in the associate family of the Enneper surface. For a sufficiently thin shell, we can assume the Kirchhoff-Love hypothesis to write the shape as \cite{love1944treatise}
\begin{equation}
\label{KL hypothesis}
\mathbf{X}_{\text{3D}}(r,\phi,s; \theta + \bar{\theta}) = \mathbf{X}_{\text{2D}}(r,\phi; \theta + \bar{\theta}) + s\hat{\mathbf{N}}(r,\phi; \theta)
\end{equation}
where $\mathbf{X}_{\text{2D}}(r,\phi; \theta(t))$ is the shell midsurface and assumed to be of the form of equation (\ref{Pinned surface parameterization}) with unit normal vector as in equation (\ref{Normal vector}), and $s \in [-\frac{h}{2},\frac{h}{2}]$, where $h$ is the shell thickness, parametrizes the direction orthogonal to the midsurface. The constraint becomes $\boldsymbol{\chi} - \mathbf{X}_{\text{3D}}(\theta + \bar{\theta}) = 0$, which  we include in the functional (\ref{action 3D KV}) through a Lagrange multiplier, $\lambda$. Variations with respect to the embedding and the Lagrange multiplier result in the equations
\begin{eqnarray}
\label{Normal force}
D_i\left(\sigma_{\text{3D}}^{ij}\partial_j \boldsymbol{\chi}\right) - \rho \partial_{tt}\boldsymbol{\chi} &=& \lambda, \\
\label{Constraint KV}
\boldsymbol{\chi} &=& \mathbf{X}_{\text{3D}}\left(\theta + \bar{\theta}\right).
\end{eqnarray}
where $D_i$ denotes the covariant derivative in the $x^i$ coordinate with respect to the prescribed metric. The Lagrange multiplier enforcing the constraint is orthogonal to the constraint set itself, whose tangential variations are given by $\delta \boldsymbol{\chi} = \partial_\theta \boldsymbol{\chi} ~ \delta \theta$. Thus, the equation of motion on the constraint set reads
\begin{equation}
\left[D_i\left(\sigma_{\text{3D}}^{ij}\partial_j \mathbf{X}_{\text{3D}}\right) - \rho \ \partial_{tt}\mathbf{X}_{\text{3D}}\right] \cdot \partial_\theta \mathbf{X}_{\text{3D}} = 0.
\label{3D constraint equation}
\end{equation}
Equation (\ref{3D constraint equation}) is a differential equation for $\theta(t)$ after prescribing a constitutive equation relating the stress to the strain and integrating over the volume of the solid. 

The 3D induced metric is
\begin{eqnarray}
    g_{ss}^{\text{3D}} &=& 1, ~  g_{sj}^{\text{3D}} = 0,\\
    g_{ij}^{\text{3D}} &=&  g_{ij}^{\text{2D}} - 2 s b_{ij} + s^2 b_{ik} b_j^k,
\end{eqnarray}
where $g_{ij}^{\text{2D}} = \partial_i \mathbf{X}_{\text{2D}} \cdot \partial_j \mathbf{X}_{\text{2D}}$ and $b_{ij}$ are the metric and second fundamental form of the midsurface, respectively. Although we have suppressed the arguments, the 3D induced metric depends on $\theta$ through the second fundamental form. We choose the prescribed 3D metric as $\bar{g}^{\text{3D}} = g^{\text{3D}}\big \rvert_{\theta = \bar{\theta}}$, where $\bar{\theta}$ is the Bonnet angle of the preferred surface in the associate family.

\subsection{Kelvin-Voigt model of viscoelasticity}

We consider a three-dimensional Kelvin-Voigt model for the total stress tensor \cite{christensen2013theory}
\begin{equation}
\sigma^{ij}_{\text{Total}} = \sigma^{ij}_{\text{elastic}} + \sigma^{ij}_{\text{dissipation}} = A^{ijkl}\gamma_{kl}^{\text{3D}} + \eta^{ijkl}\partial_t\gamma_{kl}^{\text{3D}}.
\label{kelvin-Voigt}
\end{equation}
The elastic stress, $\sigma_{\text{elastic}}$, follows Hookean elasticity with an elasticity tensor, $A^{ijkl}$, characterized by the Lam\'e parameters $\lambda$ and $\mu$ of the solid,
\begin{equation}
A^{ijkl} = \lambda \bar{g}^{ij}_{\text{3D}}\bar{g}^{kl}_{\text{3D}} + 2\mu \bar{g}^{ik}_{\text{3D}}\bar{g}^{jl}_{\text{3D}}.
\label{Hookean stress}
\end{equation}
The dissipative stress, $\sigma_{\text{dissipation}}$, is proportional to the strain rate tensor, $\partial_t \gamma^{\text{3D}}$, with a viscosity tensor written in terms of two viscosity coefficients, $\alpha$ and $\tau$,
\begin{equation}
\eta^{ijkl} = \alpha \bar{g}^{ij}_{3D}\bar{g}^{kl}_{3D} + 2\tau\bar{g}^{ik}_{3D}\bar{g}^{jl}_{3D}.
\label{viscosity tensor}
\end{equation}

Assuming the shell thickness is smaller than the characteristic radius of curvature of the midsurface, $h << \mathrm{R}$, we insert these constitutive relations into equation (\ref{3D constraint equation}), expand all terms in powers of the thickness up to order $\mathcal{O}\left(h^5\right)$ as in \cite{efrati2009elastic}, and integrate each term over the volume of the shell to obtain an equation of motion for $\theta(t)$
\begin{equation}
\mathrm{I}\ddot{\theta}(t) + 2\tau h^3 \dot{\theta}(t) +  h^3 \bar{k} \sin \theta(t)  = 0,
\label{equation of motion pendulum}
\end{equation}
where $\bar{k} = 2\mu$ \cite{efrati2009elastic}. We identify the dissipation coefficient as $\beta = \tau h^3$, with a relaxation timescale given by $t_{\text{relax}} = \mathrm{I}/(\tau h^3)$. The effective moment of inertia is given by
\begin{equation}
\mathrm{I} \approx c^2 h \mathrm{R}^4\left(1.1005 \left(\frac{h}{\mathrm{R}}\right)^2 + 12.2165 c^2 \right)\rho
\end{equation}
which, indeed, has units of mass times squared length as in other rotational systems.

\subsection{Mapping to a pendulum}

The energy-conserving part of the equation of motion (\ref{equation of motion pendulum}) can be obtained from extremizing the Lagrangian
\begin{equation}
\mathcal{L} = \frac{\mathrm{I}}{2}\left(\partial_t \theta\right)^2 - h^3 \bar{k}\left[1 - \cos \theta(t)\right],
\label{Lagrangian}
\end{equation}
from which we obtain the material parameters of a pendulum equivalent to our system by comparing it to the Lagrangian in polar coordinates of a pendulum \cite{llandau60:mech}
\begin{equation}
\mathcal{L}_\text{pendulum} = \frac{ml^2}{2}\left(\partial_t \psi\right)^2 - mgl\left(1 - \cos \psi\right),
\label{Lagrangian pendulum}
\end{equation}
where $l$ is the length of the pendulum, $m$ is the mass of pendulum, $g$ is the gravitational constant, and $\psi$ is the angle with respect to the vertical. Thus,
\begin{equation}
\psi = \theta, \quad m l^2 = \mathrm{I}, \quad m g l = h^3 \bar{k}.
\label{pendulum parameters}
\end{equation}
The mass and length for fixed $g$ are given by
\begin{equation}
l = \frac{g \mathrm{I} }{h^3 \bar{k}}, \quad m =  \frac{\left(h^3 \bar{k}\right)^2}{g^2\mathrm{I}}
\end{equation}
To derive scaling laws for the mass and length of the pendulum in terms of the thickness and radius of curvature of the sheet, we further assume a constant density $\rho = M/V$, where $M$ is the mass and $V$ the volume of the sheet. Dropping terms of order $h^2/\mathrm{R}^2$, we obtain $V \propto c^2 h \mathrm{R}^2$ and $\mathrm{I} \propto c^4 h \mathrm{R}^4 M/V \propto c^2 \mathrm{R}^2 M$. Thus,
\begin{equation}
    l \propto \frac{c^2 \mathrm{R}^2 M}{h^3 \bar{k}}, \quad m \propto \frac{h^6 \bar{k}^2}{c^2 \mathrm{R}^2 M}.
\end{equation}

\subsection{Relaxation dynamics}

\begin{figure}[H]
    \centering
    \includegraphics[width=1\linewidth]{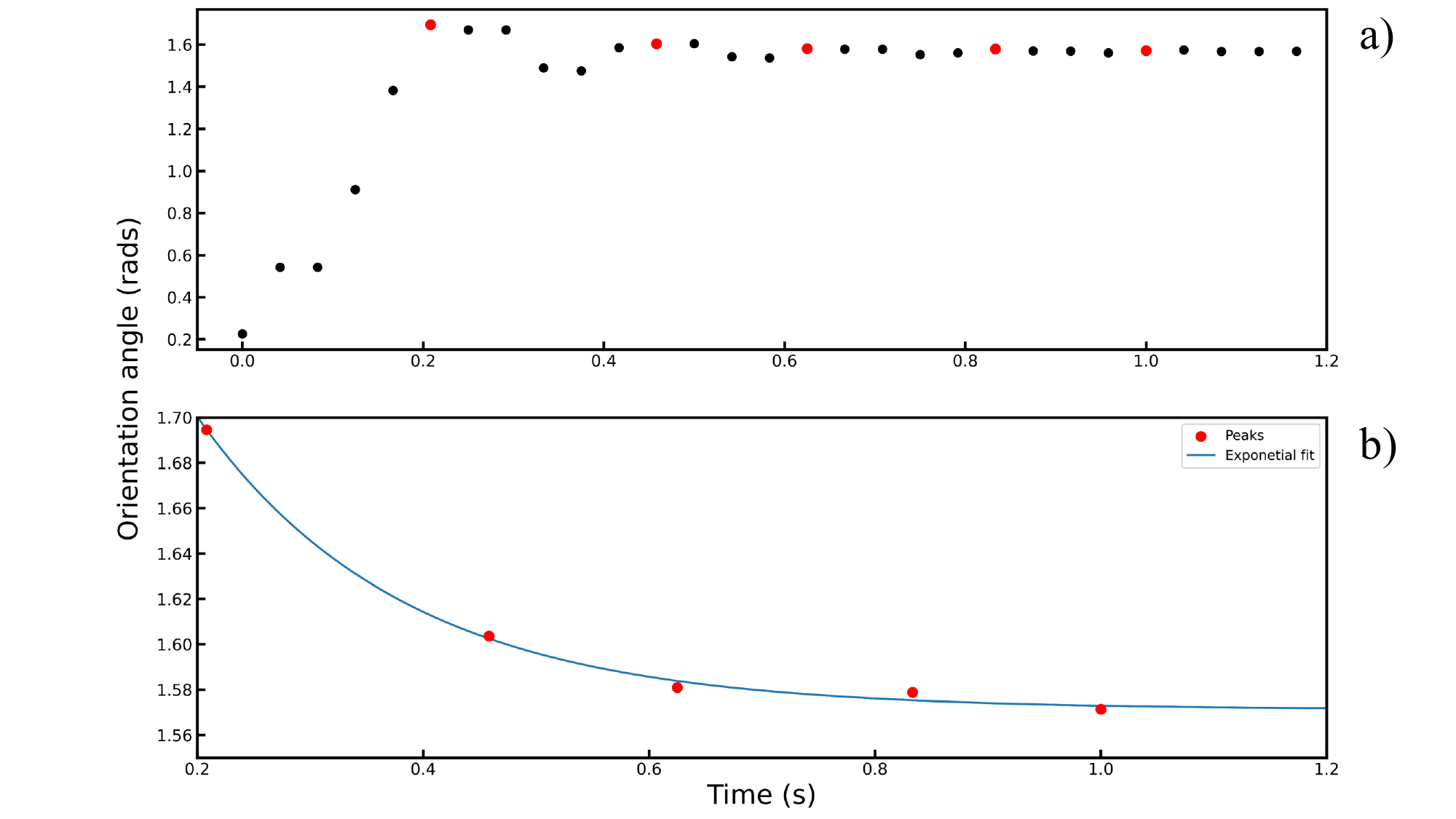}
    \caption{Orientation angle 120 hours after fabrication. a) Peaks of the oscillations (red dots) as the orientation relaxes. b) Exponential fit to peaks of the form $\delta \theta = A \exp\left[{\left(t - t_0\right)b}\right] + c/2$, from which we extract the relaxation time, $t_\text{relax} = 1/b$.}
    \label{fig:S5}
\end{figure}

\begin{figure}[H]
    \centering
    \includegraphics[width=0.8\linewidth]{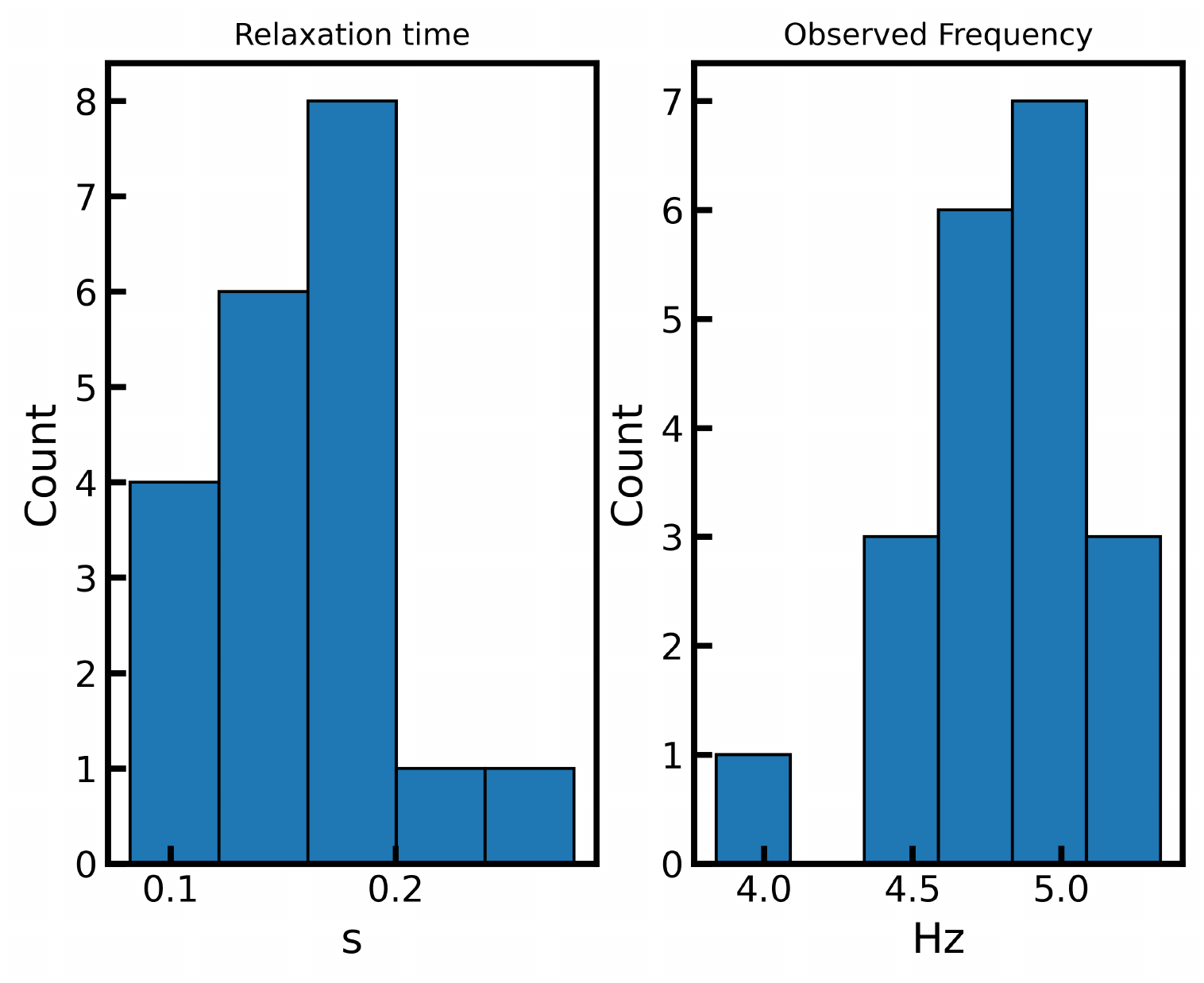}
    \caption{Extracted relaxation times and observed frequencies. Assuming an observed frequency $\approx 5$ Hz and a relaxation time $\approx 1/6$ s, we estimate a natural frequency $f_0 \approx 6$ Hz.}
    \label{fig:S6}
\end{figure}

Fig.~\ref{fig:S5} shows the experimental orientation angle 120 hours after fabrication when the plate is in the underdamped limit. We estimated the relaxation time by extracting the peaks during the oscillations to equilibrium and fitting an exponential to the data points. The equilibrium configurations sat at approximately $\pi/2$ because MATLAB's "regionprops" function measures the angle relative to the horizontal. Fig.~\ref{fig:S6} shows histograms of the estimated values. See \cite{Note1} for histograms data, and the Mathematica file with analysis.


\subsection{Overdamped-underdamped transition and sheet aging}
To study the transition from overdamped to underdamped dynamics, we assume a prescribed 3D metric with the same functional form as before, but $\bar{b} = \delta b(\bar{\theta})$, where $b(\bar{\theta})$ is the second fundamental form corresponding to a member of the associate family of the Enneper minimmal surface. The dimensionless parameter $\delta$ interpolates between a non-Euclidean plate at $\delta  = 0$ and a sheet at $\delta = 1$. Thus, it is an effective description of the aging dynamics. 

While experiments show that $\delta(t)$ depends on time, we approximate it as a constant, $\delta(t) = \delta$. This is equivalent to approximating the strain rate tensor as
\begin{equation}
    \partial_t \gamma =\frac{1}{2}\partial_\theta g^\text{3D} \dot{\theta} - \frac{1}{2}\partial_\delta \bar{g}^\text{3D} \dot{\delta} \approx \frac{1}{2}\partial_\theta g^\text{3D} \dot{\theta},
\end{equation}
which is justified since the prescribed metric evolved on the timescale of days due to aging, while the induced metric evolved on the timescale of seconds along the Bonnet isometry. Employing the Kelvin-Voigt model and dimensionality reduction once again, we arrive at Eq.~(\ref{equation of motion pendulum}) with renormalized coefficients $\bar{k} \mapsto \delta \bar{k}$ and $\mathrm{I} \mapsto \mathrm{I} + \mathrm{I}_\delta \delta^2$, where $\mathrm{I}_\delta$ is given by 
\begin{equation}
    \mathrm{I}_\delta \approx -0.375c^2 h^3 \mathrm{R}^3 \rho.
\end{equation}
Importantly, the shear viscosity $\tau$ doesn't change as $\delta$ varies. Thus, $\omega_0^2 \to 0$ but $1/t_\text{relax}^2$ remains a finite non-zero value as $\delta \to 0$. The observed angular velocity of small oscillations, $\omega_D$, requires the difference $\omega_0^2 - 1/t_\text{relax}^2$ to be positive. For sufficiently small $\delta$, however, this difference must be negative and $\delta \theta$ describes exponentially decaying behavior instead.

\section{Exciting the Bonnet isometry of Enneper's minimal surface}

\subsection{Translational modes of the sheet}

To test if the Bonnet isometry can be excited using translations, we consider adding a dynamical degree of freedom for translational modes of the center of mass of the sheet and write the midsurface shape as
\begin{equation}
    \mathbf{X}_{\text{2D}} = \mathbf{R}_\theta[\bold{\hat{z}}]\left(\cos \left[\theta + \bar{\theta}\right] ~ \mathbf{X}_0 + \sin \left[\theta + \bar{\theta}\right] ~ \mathbf{X}_C\right) + z(t) ~ \bold{\hat{z}},
\end{equation}
where the first term is the deformation due to the Bonnet isometry and the second term is the translational mode along $\bold{\hat{z}}$. A similar analysis works for modes along $\bold{\hat{x}}$ and $\bold{\hat{y}}$.

Assuming the Kirchkoff-love hypothesis, we can compute the Lagrangian in the action (\ref{action 3D KV}) for this deformation. Since there is no elastic work done by the translational mode, $\theta(t)$ and $z(t)$ can only couple through the kinetic energy. However, integration over the angular coordinate shows that
\begin{eqnarray}
    S[\mathbf{X}_{\text{2D}} + s\hat{\mathbf{N}}] &=& \int \mathrm{dt}\mathrm{dV} ~ \frac{\rho}{2}\left(\partial_t\left[\mathbf{X}_{\text{2D}} + s\hat{\mathbf{N}}\right]\right)^2, \nonumber\\
    &=& \int \mathrm{dt}\mathrm{dr}\mathrm{ds} ~  \frac{\sqrt{\det \bar{g}}\rho}{2}\left[2\pi \left(\partial_t z\right)^2 + \frac{\pi r^2}{9\mathrm{R}^4}\left(c^2\left[8r^4 + 9 r^2 \mathrm{R}^2\right] + \frac{72\mathrm{R}^6s^2}{\left(r^2 + \mathrm{R}^2\right)^2}\right)\left(\partial_t \theta\right)^2\right],
\end{eqnarray}
since the volume element is independent of $\phi$. Thus, the translational mode and the Bonnet angle are decoupled. A similar result is obtained for the other two translational modes.

\subsection{Magnetic driving}

\begin{figure}[h!]
    \centering
    \includegraphics[width=0.6\linewidth]{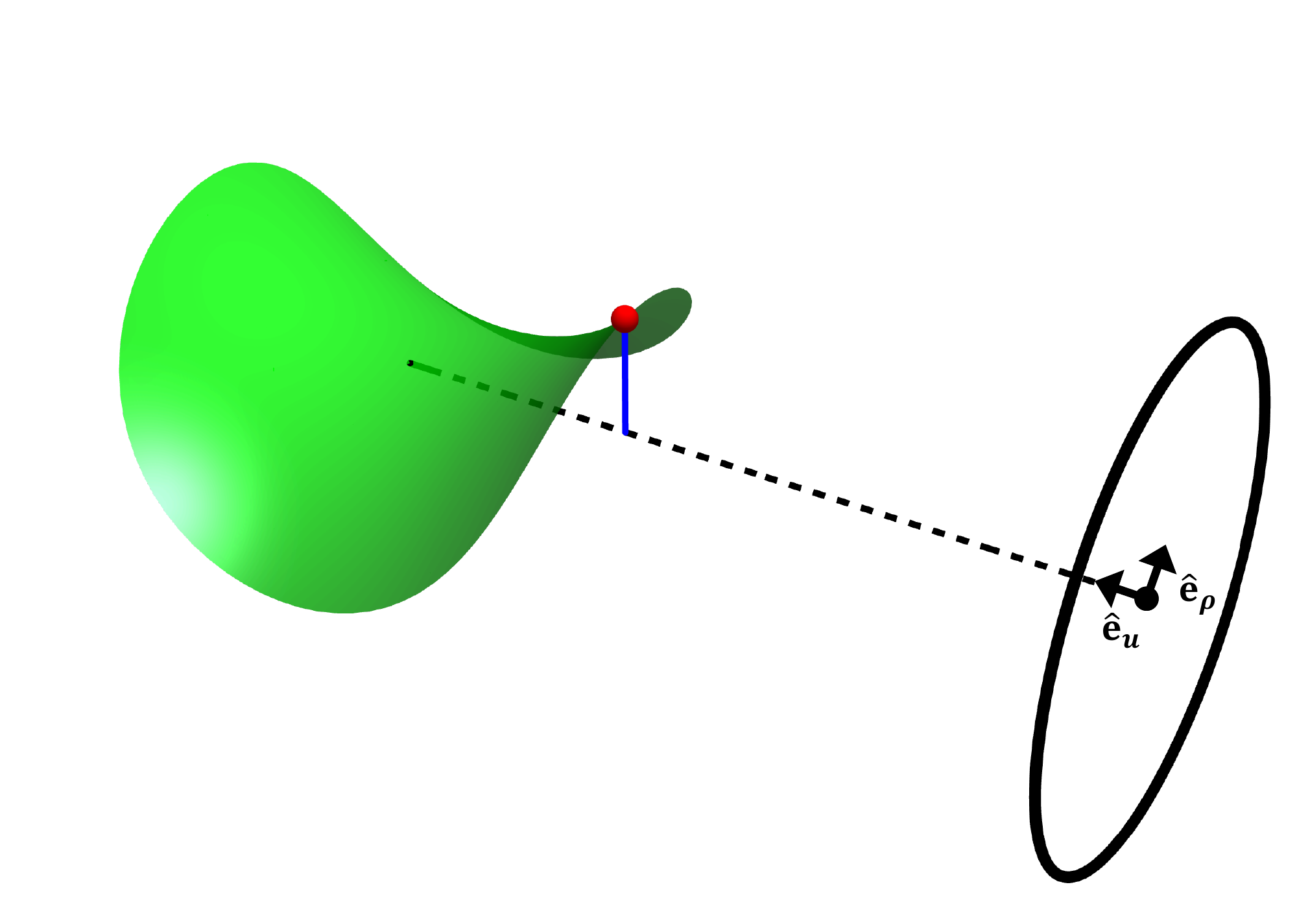}
    \caption{Schematic of the magnetic driving of the Enneper minimal surface. The spinning magnet is modeled as a current $\bold{J} = j \cos(\omega t) \hat{\bold{e}}_\psi$ going around a circular wire (black circle). The wire's center, plate's center, and magnet on the surface (red dot) are collinear. The distance from the center of the wire to the edge of the plate where the magnet is (dotted line) is $d$. The magnet on the surface sits at a height $z \propto \mathrm{R}\cos(\theta + \bar{\theta} + 2\phi_0)$ (blue line) relative to the dotted line, where $\phi_0$ is the polar angle coordinate of the magnet.}
    \label{fig:4}
\end{figure}

To model the magnetic driving in our experiments, we assume the magnet attached to the Enneper surface is a magnetic dipole, sitting at $p_0 = \mathbf{X}_{\text{3D}}\left(r_0, \phi_0, s_0; \bar{\theta}\right)$, whose three-dimensional dipole moment $\bold{m}$ is interacting with an external magnetic force $\bold{B}$ produced by another rotating magnetic dipole away from the surface. We assume $r_0 = \mathrm{R}$ and $s_0 = h/2$ for simplicity. Since the magnet was considerably smaller than the shell, we assume that the dipole moment is aligned with the unit normal of the midsurface, $\bold{m} = m_s \hat{\bold{N}}$, where $m_s$ is a constant magnetic dipole.
We model the rotating magnet as an Amperian circular loop of radius $r_m$, whose center is located at a distance $d$ from the center of the Enneper plate. In our experiments, the magnet on the surface, the center of the spinning magnet, and the center of the plate were aligned at rest. Note, however, that the magnet on the plate sat at a height $\propto \mathrm{R} \cos \left(\bar{\theta} + 2\phi_0\right)$ relative to the two centers (Fig.~\ref{fig:4}). We consider a potential energy of the form \cite{griffiths2023introduction, jackson2021classical}
\begin{equation}
\label{magnetic energy}
\mathcal{U}_m = - \int \mathrm{dV} \mathrm{dt} ~ \frac{\delta(r - r_0) \delta(\phi - \phi_0)\delta(s - s_0)}{\sqrt{\det \bar{g}}} ~ \bold{m}\cdot \bold{B} = -\int \mathrm{dt} ~ \left(\bold{m}\cdot \bold{B}\right) \rvert_{\left(r = r_0, \phi = \phi_0, s = s_0\right)}
\end{equation}
where the integration is over the volume given by the Kirchhoff-Love hypothesis (\ref{KL hypothesis}). 

To compute the magnetic field, it is convenient to introduce cylindrical coordinates, $(\rho, \psi, u)$, about the center of the loop and relative to the edge of the plate (the intersection between the blue line and the dotted line in Fig.~\ref{fig:4}) as $x_c = \rho \cos \psi, \ y_c = u + d, \ z_c = \rho \sin \psi$. We consider an alternating current along the loop, $\bold{J} = j \cos(\omega t) \hat{\bold{e}}_\psi$, where $\omega$ is the frequency of the rotating magnet, and $j$ is a constant current per unit length. When the product of the applied frequency and distance from the loop is small compared to the speed of light, the magnetic field is well approximated by \cite{griffiths2023introduction, jackson2021classical, gonzalez2020analytical}
\begin{eqnarray}
\label{Magnetic field}
\bold{B} = \frac{\mu_0 j \cos \omega t}{2\pi \sqrt{(r_m + \rho)^2 + u^2}}\left(\frac{u}{\rho}\left[\left(\frac{r_m^2 + \rho^2 + u^2}{(\rho - r_m)^2 + u^2}\right)\mathrm{E}(m) - \mathrm{K}(m)\right]\hat{\bold{e}}_\rho -  \left[\left(\frac{\rho^2 - r_m^2 + u^2}{(\rho - r_m)^2 + u^2}\right)\mathrm{E}(m) - \mathrm{K}(m)\right]\hat{\bold{e}}_u\right), \nonumber\\
\end{eqnarray}
where $\hat{\bold{e}}_\rho$, $\hat{\bold{e}}_\psi$,$\hat{\bold{e}}_u$ are the cylindrical unit vectors, $\mu_0$ is the vacuum magnetic permeability, and $\mathrm{E}(m)$ and $\mathrm{K}(m)$ are the complete elliptic integrals of first and second kind with parameter $m = 4r_m \rho/\left[(r_m + \rho)^2 + u^2\right]$. The position vector $\boldsymbol{r}$ from the center of the loop to the magnet on the plate can be written in terms of the embedding as 
\begin{eqnarray}
    \boldsymbol{r} &=& -\left[d + \mathbf{X}_{\text{3D}}\left(p_0; \bar{\theta}\right)\cdot \hat{\bold{y}}\right] \hat{\bold{e}}_u + \mathbf{X}_{\text{3D}}\left(p_0; \theta + \bar{\theta}\right),\\
    &=& \left[\Delta \mathbf{X}_{\text{3D}}\left(p_0 \right) \cdot \hat{\bold{y}} - d\right]\hat{\bold{e}}_u + \left[\mathbf{X}_{\text{3D}}\left(p_0; \theta + \bar{\theta}\right) \cdot \hat{\bold{x}} ~ \hat{\bold{x}} + \mathbf{X}_{\text{3D}}\left(p_0; \theta + \bar{\theta}\right) \cdot \hat{\bold{z}} ~ \hat{\bold{z}}\right],
\end{eqnarray}
where we have decomposed the embedding in its Cartesian components, identified $\hat{\bold{e}}_u = \hat{\bold{y}}$, defined the displacement $\Delta \mathbf{X}_{\text{3D}}\left(p_0 \right) = \mathbf{X}_{\text{3D}}\left(p_0; \theta + \bar{\theta}\right) - \mathbf{X}_{\text{3D}}\left(p_0;  \bar{\theta}\right)$, and set $p_0 = (r_0, \phi_0, s_0)$ for brevity. Thus, the cylindrical coordinates of the magnet are
\begin{eqnarray}
    u &=& \Delta \mathbf{X}_{\text{3D}}\left(p_0 \right) \cdot \hat{\bold{y}} - d,\\
    \rho^2 &=& \left[\mathbf{X}_{\text{3D}}\left(p_0; \theta + \bar{\theta}\right) \cdot \hat{\bold{x}}\right]^2 + \left[\mathbf{X}_{\text{3D}}\left(p_0; \theta + \bar{\theta}\right) \cdot \hat{\bold{z}}\right]^2,\\
    \hat{\bold{e}}_\rho &=& \cos \psi \hat{\bold{x}} + \sin \psi \hat{\bold{z}} = \frac{\mathbf{X}_{\text{3D}}\left(p_0; \theta + \bar{\theta}\right) \cdot \hat{\bold{x}}}{\rho} ~ \hat{\bold{x}} + \frac{\mathbf{X}_{\text{3D}}\left(p_0; \theta + \bar{\theta}\right) \cdot \hat{\bold{z}}}{\rho} ~ \hat{\bold{z}}.
\end{eqnarray}

The natural length parameter of the embedding is the radius of curvature $\mathbf{X}_{\text{3D}} \sim \mathrm{R}$ up to terms of order $h/\mathrm{R}$. Thus, $u \sim \mathrm{R}$ up to terms of order $d/\mathrm{R}$ and $\rho^2 \propto \mathrm{R}^2$. Since the radius of the spinning magnet $r_m$ was much smaller than $\mathrm{R}$ in our experiments, $r_m/\mathrm{R} \approx 0.16$, we Taylor expand Eq.~(\ref{Magnetic field}) to leading order in $r_m/\mathrm{R}$. In particular, the elliptic integrals are well approximated by their small $m$ expansions since $m \propto r_m/\mathrm{R}$. Keeping terms of order $d/\mathrm{R}$, we obtain 
\begin{eqnarray}
    \bold{B} &=& \frac{\mu_0 j \cos \omega t}{4}\left(\frac{\Tilde{\rho}}{\left[\Tilde{u}^2 + \Tilde{\rho}^2\right]^{3/2}}\right)\left(\frac{\Tilde{u}}{\Tilde{\rho}}\hat{\bold{e}}_\rho - \hat{\bold{e}}_u\right)\frac{r_m}{\mathrm{R}} + \mathcal{O}\left[\left(\frac{r_m}{\mathrm{R}}\right)^2\right],\\
    &=& \frac{\mu_0 j \cos \omega t}{4}\frac{\left(\Tilde{u}\hat{\bold{e}}_\rho - \Tilde{\rho}\hat{\bold{e}}_u\right)}{\left(\Tilde{u}^2 + \Tilde{\rho}^2\right)^{3/2}}\frac{r_m}{\mathrm{R}} + \mathcal{O}\left[\left(\frac{r_m}{\mathrm{R}}\right)^2\right]
\end{eqnarray}
where we have defined $\rho = \mathrm{R}\Tilde{\rho}$ and $u = \mathrm{R}\Tilde{u}$. 

\begin{figure}[H]
    \centering
    \includegraphics[width=1\linewidth]{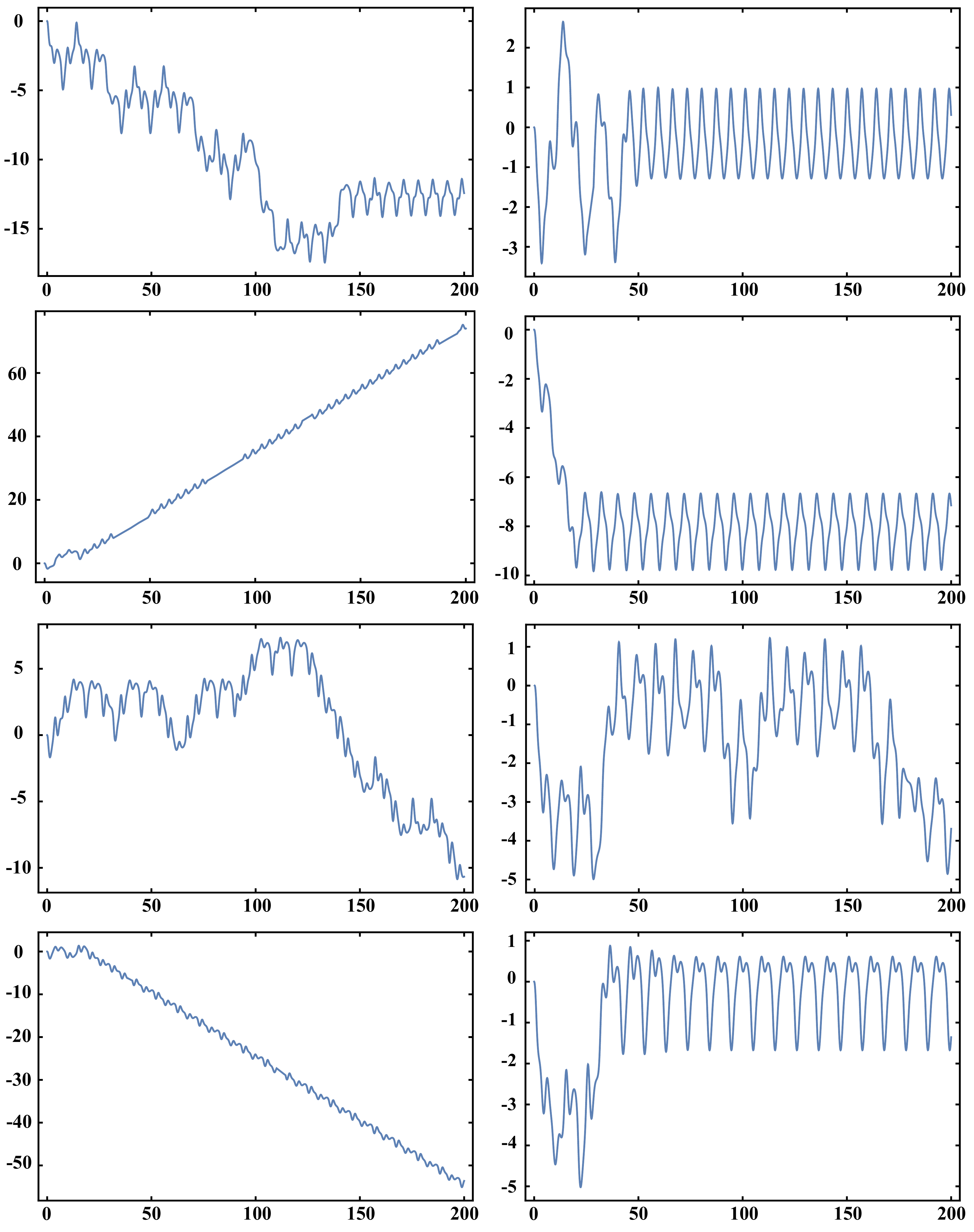}
    \caption{Numerical solutions of Eq.~(\ref{dimensionless driven pendulum}) for the orientation angle $\theta/2$ (rads) as a function of dimensionless time $\bar{t} = \omega_0 t$ for different driving frequencies (rows, from bottom to top: $3, 3.5, 3.95, 4.5$ Hz) and distances (columns, left: $5$ cm. Right: $10$ cm), with the sheet's magnet at $\phi_0 = 0$ and $r_0 = R$, showing spinning modes at close distances and bounded oscillations at long distances over long time. We have set the driving coefficient to 10.}
    \label{fig:Bonnet numerics}
\end{figure}

Adding the magnetic interaction in Eq.~(\ref{magnetic energy}) to the Lagrangian (\ref{Lagrangian}), as a potential energy with the magnetic field up to leading order in $r_m/\mathrm{R}$, the equation of motion is
\begin{equation}
\label{Driven pendulum}
\mathrm{I}\ddot{\theta}(t) + 2\tau h^3 \dot{\theta}(t) +  h^3 \bar{k} \sin\theta(t)  -\frac{r_m}{\mathrm{R}} \frac{m_s \mu_0 j \cos \omega t}{4}\mathcal{N}[\theta]= 0,
\end{equation}
where the nonlinearity $\mathcal{N}[\theta]$ is given by
\begin{equation}
\label{nonlinearity magnetic}
\mathcal{N}[\theta] =  \partial_\theta\left[ \hat{\mathbf{N}}\cdot \frac{\left(\Tilde{u}\hat{\bold{e}}_\rho - \Tilde{\rho}\hat{\bold{e}}_u\right)}{\left(\Tilde{u}^2 + \Tilde{\rho}^2\right)^{3/2}}\bigg \rvert_{p_0}\right],
\end{equation}
with the component of the magnetic field evaluated at $p_0$. It is more convenient to work with Eq.~(\ref{Driven pendulum}) after introducing a dimensionless time $\bar{t} = \omega_0 t$ in units of the natural angular velocity
\begin{equation}
\label{dimensionless driven pendulum}
\ddot{\theta}(t) + \frac{2}{t_\text{relax}\omega_0} \dot{\theta}(t) + \sin\theta(t)  - \frac{m_s \mu_0 j r_m\cos \left(\frac{\omega}{\omega_0} \bar{t}\right)}{4\mathrm{R}\mathrm{I}\omega_0^2}\mathcal{N}[\theta]= 0.
\end{equation}
In particular, $\omega/\omega_0 = f/f_0 \approx f/6$ and $2/(t_\text{relax}\omega_0) \approx 0.4$ with our previous estimates from the free dynamics. Here we use a slightly more precise approximation of the moment of inertia, $\mathrm{I} \propto c^2\mathrm{R}^2 \mathrm{M}/(2\pi)$. The coefficient of the driving term is approximately $m_s \mu_0 J/ \left(4c^2 \mathrm{M}\mathrm{R}^3\omega_0^2\right)$, where $J$ is the total current along the Amperian loop. The sheet has parameters $\mathrm{M} \approx 2$ g, $\mathrm{R} \approx 10$ cm, and $c \approx 0.7$, the neodymium magnet has magnetic moment $m_s \approx 30$ A~m$^2$ and we estimate the total current as $J \approx 1-2$ A. Thus, a conservative estimate of the driving coefficient is $5 - 10$. Fig.~\ref{fig:Bonnet numerics} shows numerical solutions of Eq.~(\ref{Driven pendulum}) using these estimates as a function of the driving frequency and the distance between magnets, qualitatively in agreement with the experimental data in Fig~\ref{fig:Magnetic driving data}.

\subsection{Small deformations away from the preferred configuration}
To study the transition from spinning to small oscillations in our experimental data, we consider a small deformation, $\theta = \delta \theta$, about the reference configuration and linearize Eq.~(\ref{dimensionless driven pendulum}) to obtain
\begin{equation}
\ddot{\delta\theta} + \frac{2}{t_\text{relax}\omega_0} \dot{\delta\theta} + \left[1 - \frac{r_m}{\mathrm{R}}\frac{m_s \mu_0 j}{4\mathrm{I}\omega_0^2}\partial_\theta\mathcal{N}[0]\cos \left(\frac{\omega}{\omega_0}\bar{t}\right)\right]\delta \theta = \frac{r_m}{\mathrm{R}} \frac{m_s \mu_0 j }{4\mathrm{I}\omega_0^2}\mathcal{N}[0]\cos \left(\frac{\omega}{\omega_0} \bar{t}\right)
\label{linearized pendulum}
\end{equation}
where the derivative of the nonlinearity is computed before evaluating at $\theta = 0$. We identify two small dimensionless numbers in the equation: $r_m/\mathrm{R} \approx 0.16$ and $1/(t_\text{relax}\omega_0) \approx 0.2$ from previous estimates. Since both are about the same size, we choose to work with $\epsilon = r_m/\mathrm{R}$ and write $1/(t_\text{relax}\omega_0) = \epsilon \mu$ with $\mu = \mathcal{O}(1)$. Thus, the small deformation is controlled by a forced Mathieu equation in the limit of weak damping and forcing
\begin{equation}
    \ddot{\delta\theta} + 2\epsilon \mu \dot{\delta\theta} + \left[1 + \epsilon \gamma\cos \left(\frac{\omega}{\omega_0}\bar{t}\right)\right]\delta \theta = \epsilon F \cos \left(\frac{\omega}{\omega_0} \bar{t}\right)
    \label{Papers Mathieu}
\end{equation}
as studied in \cite{ramakrishnan2012resonances, ramakrishnan2022primary} with undamped modal frequency $1$, phase angle $\pi/2$ and parameters 
\begin{eqnarray}
    \Omega = \frac{\omega}{\omega_0}, ~ \mu = \frac{\mathrm{R}}{r_m t_\text{relax}\omega_0}, ~ \gamma = -\frac{m_s \mu_0 j}{4\mathrm{I}\omega_0^2}\partial_\theta\mathcal{N}[0], ~ F = \frac{m_s \mu_0 j }{4\mathrm{I}\omega_0^2}\mathcal{N}[0].
\end{eqnarray}

We summarize the analysis in \cite{ramakrishnan2012resonances, ramakrishnan2022primary}, with an emphasis on stability properties. To second order in $\epsilon$, solutions to Eq.~(\ref{Papers Mathieu}) exhibit three resonances: the primary resonance $\omega/\omega_0 \approx 1$ and a subharmonic resonance $\omega/\omega_0 \approx 2$ at first order in $\epsilon$, and a superharmonic resonance $\omega/\omega_0 \approx 1/2$ at second order in $\epsilon$. Since linearly unstable modes grow exponentially, but are stabilized by nonlinearities at later times, we expect spinning-like modes to occur near these resonances. Motivated by experimental data on spinning modes, we focus on the subharmonic resonance. Solutions to Eq.~(\ref{Papers Mathieu}) are linear combinations of a homogeneous solution, satisfying the unforced Mathieu equation, and a particular solution that includes the external forcing term. The latter produces a bounded oscillatory solution, while the homogeneous solution becomes unstable when 
\begin{equation}
\label{Instability boundary}
0 = 4\left(\mu^2 + \sigma^2\right) - \frac{\gamma^2}{4} + \mathcal{O}\left(\epsilon\right),
\end{equation}
where $\sigma$ is a detuning parameter, $\omega = \omega_0 + \epsilon \sigma$. 

Eq.~(\ref{Instability boundary}) defines instability boundaries, $\sigma(\gamma)$ depending on the distance $d$ and sheet magnet's position $\phi_0$ through the nonlinearity $\mathcal{N}$, $\gamma(d, \phi_0)$. The boundaries are given by
\begin{equation}
    \sigma_\pm = \pm \sqrt{\frac{\gamma^2}{16} - \mu^2}.
    \label{Boundary curves}
\end{equation}
The modes are unstable within the region bounded by the boundaries. The gap between the boundaries is given by $2\sqrt{\gamma^2 \epsilon^2/16 - \mu^2 \epsilon^2}$, and the region ends at distances that satisfy $\gamma/4 = \pm \mu$, after which the square root becomes complex. 

\begin{figure}[h!]
    \centering
    \includegraphics[width=1\linewidth]{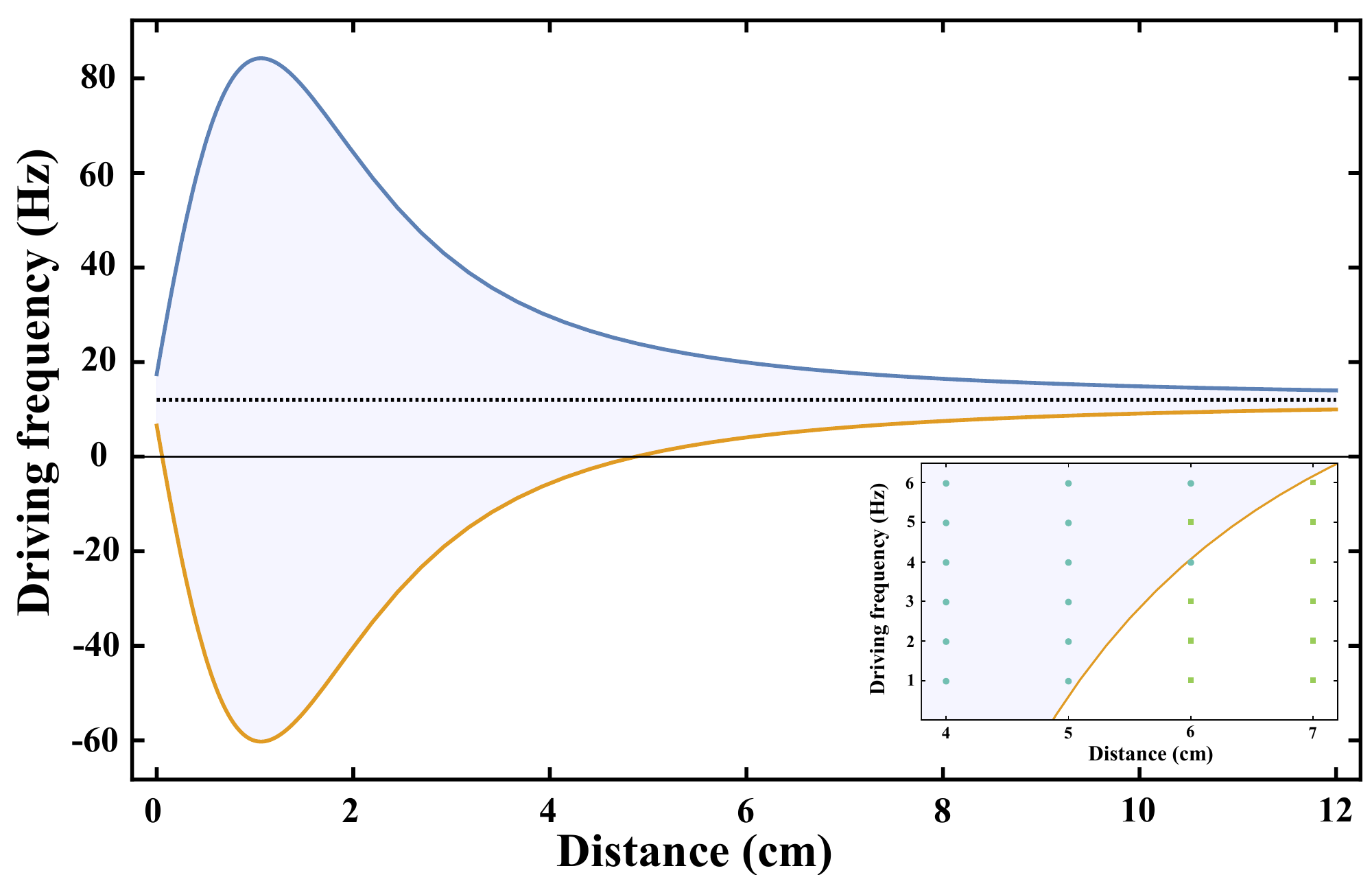}
    \caption{Predicted unstable region for the parametric resonance of the forced Mathieu Eq.~(\ref{Papers Mathieu}) using estimated parameters from the previous section and sheet's magnet placement $\phi_0 = 0$. We have used $\bar{\theta} = 1.37$, which gave the best qualitative agreement with experiments. The dotted line is at $2f_0 = 12$ Hz. The system is linearly unstable in the shaded region. Boundary curves (blue and yellow) are given by $f = 2f_0 + \epsilon \sigma/(2\pi)$, with $\sigma$ fixed by Eq.~(\ref{Boundary curves}). \emph{Inset}: Zoom-in to the parameter range in experiments of the aged sheet with data points superimposed. Blue disks: spinning mode. Green squares: bounded-
oscillation mode.}
    \label{fig:S12}
\end{figure}

Fig.~\ref{fig:S12} shows the expected unstable region around $2f_0$ with the parameters estimated from the previous section, where we see qualitative agreement with the unstable modes observed in experiments. Since $\mu \propto 1/\omega_0$ and $\gamma \propto 1/\omega_0^2$, the unstable region grows as $\omega_0 \to 0$. Hence, fresh sheets are expected to exhibit spinning modes across a wider range of parameters, even as the natural frequency $f_0$ vanishes, in agreement with Fig.~\ref{fig:aging dependence}.

\pagebreak
\bibliography{sn-bibliography}